\documentclass[lettersize,journal]{IEEEtran}
\usepackage{amsmath,amsfonts}
\usepackage{algorithmic}
\usepackage{algorithm}
\usepackage{array}
\usepackage{graphicx}
\usepackage[caption=false,font=footnotesize]{subfig}
\usepackage{textcomp}
\usepackage{stfloats}
\usepackage{url}
\usepackage{verbatim}

\usepackage{booktabs}
\usepackage{siunitx}
\usepackage{mathtools}
\usepackage{cite}

\usepackage[acronym]{glossaries}

\newacronym{csi}{CSI}{channel state information}
\newacronym{cir}{CIR}{channel impulse response}
\newacronym{cfr}{CFR}{channel frequency response}
\newacronym{stft}{STFT}{short-time Fourier transform}
\newacronym{tdl}{TDL}{tapped delay line}
\newacronym{idft}{IDFT}{inverse discrete Fourier transform}
\newacronym{ota}{OTA}{over-the-air}
\newacronym{los}{LoS}{line-of-sight}
\newacronym{rssi}{RSSI}{received signal strength indicator}
\newacronym{cots}{COTS}{commercial off-the-shelf}
\newacronym{pca}{PCA}{principal component analysis}
\newacronym{ofdm}{OFDM}{orthogonal frequency division multiplexing}
\newacronym{isac}{ISAC}{integrated sensing and communications}

\def\jj{{\rm j}}
\def\ee{{\rm e}}

\DeclareMathOperator*{\argmax}{arg\,max}
\DeclarePairedDelimiter\parens{\lparen}{\rparen}

\graphicspath{
{./_figs_1st/}
{./_figs_2nd/}
}

\usepackage{tikz}
\usetikzlibrary{arrows.meta,positioning,shapes.geometric,fit,calc,shapes.misc,decorations.pathmorphing}
\usepackage{xcolor}
\begin{document}
\title{Wi-Fi Radar via Over-the-Air Referencing: \\
Bridging Wi-Fi Sensing and Bistatic Radar}
\author{Koji Yamamoto,~\IEEEmembership{Senior Member,~IEEE}
%<-this % stops a space
%\thanks{This paper was produced by the IEEE Publication Technology Group. They are in Piscataway, NJ.}% <-this % stops a space
%\thanks{Manuscript received April 19, 2021; revised August 16, 2021.}}
\thanks{This work was supported in part by JSPS KAKENHI Grant Numbers JP23K24831 and JP23K26109.}%
\thanks{K.~Yamamoto is with Faculty of Information and Human Sciences, Kyoto Institute of Technology, Matsugasaki Hashikamicho, Sakyo-ku, Kyoto 606-8585, Japan (e-mail: \mbox{kyamamot@kit.ac.jp}). }
}
%\markboth{\today}%
\markboth{}%
{}
\maketitle

\begin{abstract}

Wi-Fi channel state information (CSI), which is originally acquired
for communication purposes, has recently been reused for sensing
and radar-like functionalities.
However, in practical Wi-Fi systems with independent clocks at the
transmitter and receiver, the lack of a common delay and phase reference
fundamentally precludes phase-coherent radar-like delay--Doppler analysis.
By exploiting the line-of-sight (LoS) path component, i.e., the earliest-arriving direct path,
as an over-the-air (OTA) reference for delay and phase,
we propose an OTA LoS-path referencing scheme, termed LoSRef,
that enables delay calibration and phase alignment under this practical constraint.
Unlike conventional \mbox{Wi-Fi} bistatic radar systems that rely on wired
reference signals or dedicated reference antennas, the proposed
LoSRef-based framework enables phase-coherent bistatic radar-like
operation that can be integrated into typically deployed \mbox{Wi-Fi} systems.
Through human gait and respiration experiments in indoor environments,
we demonstrate that phase-coherent channel impulse responses and corresponding delay--Doppler responses
can be obtained using only commodity \mbox{Wi-Fi} devices.
This enables physically interpretable human motion sensing,
including gait-induced range variation and respiration-induced
sub-wavelength displacement, as well as the extraction of
target-induced dynamics up to 20\,dB weaker than dominant static multipath components.

\end{abstract}


\section{Introduction}
\IEEEPARstart{A}{s}
a key application of \gls{isac}, which aims to augment wireless communication systems with sensing capabilities by leveraging existing wireless infrastructure \cite{Liu2022JSAC},
wireless LAN systems with device-free sensing capability have attracted significant attention in recent years \cite{Ma2019wifi,Tan2022IOT,Chen2023ACM}.
By exploiting ubiquitous \mbox{Wi-Fi} transmissions, numerous studies have demonstrated the feasibility of monitoring human respiration, gait, and daily activities without requiring the target to carry any dedicated device.
These capabilities have enabled a broad range of applications in healthcare, smart living environments, and ambient intelligence, while benefiting from the low deployment cost and widespread availability of commercial \mbox{Wi-Fi} networks.
%\cite{Wang2016UbiCompRespiration,Wang2017WiFall}.

Early \mbox{Wi-Fi} sensing studies relied on coarse metrics such as \gls{rssi} \cite{Moussa2009PERCOM,Seifeldin2013TMC}. 
A major turning point was the release of the Linux 802.11n \gls{csi} tool \cite{Halperin2011SIGCOMM}, which enabled researchers to access \gls{csi} from \gls{cots} IEEE 802.11n devices.
In this context, CSI refers to the complex-valued \gls{cfr}
measured across \gls{ofdm} subcarriers.
From a communication-system perspective, CSI is obtained for communication purposes such as equalization, precoding, and link adaptation, rather than being designed for sensing purposes.
Throughout this paper, the terms CSI and CFR are used interchangeably:
CSI is used when referring to channel measurement,
while CFR denotes the frequency-domain representation of the channel
and its Fourier-transform relationship to the \gls{cir},
which represents the channel in the delay domain.

This CSI acquisition capability has triggered research on CSI-based \mbox{Wi-Fi} sensing,
including motion-related sensing based on temporal variations
of the wireless channel, as summarized in surveys such as \cite{Ma2019wifi,Tan2022IOT,Chen2023ACM}.
More recently, \mbox{Wi-Fi} sensing has matured, culminating in the standardization of IEEE 802.11bf \cite{Du2025CSTO}, which defines channel measurement procedures to support sensing use cases, including range, motion, and gesture detection.

\begin{figure}[t]
 \centering
 \includegraphics[width=\linewidth]{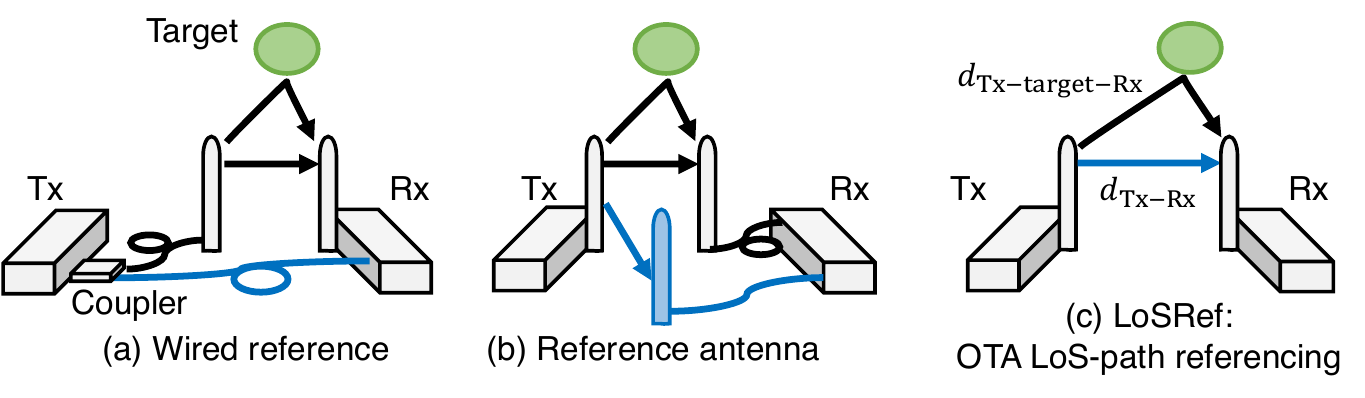}
\caption{Comparison of reference acquisition strategies.
While (a) and (b) rely on dedicated reference hardware,
the proposed OTA LoS-path referencing scheme (LoSRef) in (c) exploits the Tx--Rx
LoS path as a delay reference, eliminating the need for
auxiliary reference hardware.
Moreover, (c) is compatible with a single-port
Rx, whereas (a) and (b) require multi-port Rxs.
}
\label{fig:ref_schemes}
\end{figure}

Alongside the \mbox{Wi-Fi} sensing approaches, there exists a line of research
that exploits COTS \mbox{Wi-Fi} transmissions for radar applications \cite{Guo2008Radar,Colone2012AES}.
For clarity, we refer to methods operating without explicit reference
signals as \textit{Wi-Fi sensing}, and those employing explicit
reference signals for phase-coherent analysis as \textit{Wi-Fi radar}.
Early \mbox{Wi-Fi} radar studies typically relied on measurement-grade receivers
and external reference acquisition mechanisms, such as reference antennas
or wired signal taps, as illustrated in Fig.~\ref{fig:ref_schemes}(a) and (b),
to enable coherent delay or Doppler estimation \cite{Guo2008Radar,Chetty2012GRS,Colone2012AES,Falcone2012AES,Pastina2015VT}.
Recent works have demonstrated that reference signals can also be
exploited in fully COTS-based systems to stabilize CSI measurements, i.e., the phase of the CFR \cite{Keerativoranan2018,Song2020IOT,Qiu2023IOT,Yamamoto2025EuCAP}.
These reference-aided architectures are consistent with
bistatic radar principles and demonstrate the potential of
\mbox{Wi-Fi} signals for radar-like sensing.
However, due to their reliance on cable routing or dedicated reference antenna placement,
most prior \mbox{Wi-Fi} radar functionalities cannot be integrated into typically deployed Wi-Fi systems.

These observations reveal a fundamental gap between 
\mbox{Wi-Fi} sensing and \mbox{Wi-Fi} radar approaches.
While reference-aided \mbox{Wi-Fi} radar enables physically meaningful
delay and Doppler analysis, such architectures cannot be integrated into typical Wi-Fi systems.
Conversely, \mbox{Wi-Fi} sensing approaches that rely, for example, on CFR magnitude alone
do not require any modification to existing \mbox{Wi-Fi} systems, but lack direct access to physically
interpretable propagation parameters.

This raises a key question:
\emph{Can bistatic \mbox{Wi-Fi} radar functionality be realized using only
unmodified COTS \mbox{Wi-Fi} devices, without wired references or
dedicated reference antennas, while being integrated into typical communication-oriented \mbox{Wi-Fi} systems,
and still enabling physically meaningful delay--Doppler estimation?}

%\emph{Can bistatic \mbox{Wi-Fi} radar functionality be realized using only
%unmodified COTS \mbox{Wi-Fi} devices, without wired references or
%dedicated reference antennas, while still enabling physically meaningful delay--Doppler estimation?}

The contributions of this paper are summarized as follows.
\begin{itemize}
\item
Unlike most existing \mbox{Wi-Fi} sensing approaches that operate directly on the \gls{cfr},
which is originally acquired for communication purposes,
a \gls{cir} is constructed using \gls{cfr} obtained from
COTS IEEE 802.11ax devices with \SI{160}{\mega\hertz} bandwidth \cite{ieee80211ax}.
This is enabled by carefully designed frequency-domain preprocessing
of the raw CFR,
and provides a reliable basis for subsequent phase-coherent delay-domain processing.

\item
To the best of our knowledge, this is the first work that realizes
reference-aided bistatic \mbox{Wi-Fi} radar
that can be integrated into typically deployed \mbox{Wi-Fi} systems,
using only unmodified
COTS \mbox{Wi-Fi} devices, without relying on wired references or
dedicated reference antennas.
To this end, we propose an \gls{ota} \gls{los}-path referencing scheme (LoSRef)
that exploits the temporal stability of the transmitter--receiver (Tx--Rx) \gls{los} path to provide a common delay and phase reference in practical \mbox{Wi-Fi} systems, as illustrated in Fig.~\ref{fig:ref_schemes}(c),
thereby enabling phase-coherent CIR acquisition.

\item
The proposed LoSRef-based framework requires only a single antenna port at the Rx,
whereas conventional reference-aided bistatic \mbox{Wi-Fi} radar functionalities
typically rely on multi-port Rxs or dedicated reference channels.
This further simplifies hardware requirements and facilitates
integration into typical \mbox{Wi-Fi} systems
using unmodified COTS \mbox{Wi-Fi} devices.
%This further simplifies hardware requirements and facilitates
%drop-in deployment with COTS \mbox{Wi-Fi} devices.
     
\item
By leveraging the phase of the CIR, we demonstrate that
respiration-induced body-surface motion, as a representative example of
sub-wavelength displacement, can be directly observed
and quantitatively interpreted in indoor environments, in a manner that is
not accessible with conventional magnitude-based \mbox{Wi-Fi} sensing.
%     
%This enables physically meaningful path-length displacement information
%beyond conventional magnitude-based \mbox{Wi-Fi} sensing.

\item
We experimentally show that a delay--Doppler representation can be derived from
the CIR while preserving the sign of the Doppler frequency,
which is inherently lost in magnitude-based approaches.     
This enables discrimination between approaching and receding motions,
achieving phase-coherent bistatic radar functionality with COTS \mbox{Wi-Fi} devices.
\end{itemize}

%\medskip

While the proposed framework enables various application-level sensing tasks,
including gait analysis and respiration monitoring,
the focus of this paper is on establishing a delay-domain,
phase-coherent signal representation enabled by LoSRef.
Accordingly, the intent is not to compete with state-of-the-art
application-specific estimators or to benchmark their estimation accuracy,
but to establish a physically grounded, phase-coherent signal representation
that fundamentally differs from conventional magnitude-based Wi-Fi sensing
and upon which such estimators and applications can be built.

%to provide a physically grounded representation upon which such
%estimators and applications can be built.

This paper is organized as follows.
Section~\ref{sec:Related-Work} reviews related work on Doppler-based \mbox{Wi-Fi} sensing and reference-aided \mbox{Wi-Fi} radar.
Section~\ref{sec:CIR-Representation} discusses the reference ambiguity due to independent clocks in practical \mbox{Wi-Fi} systems.
Section~\ref{sec:time-variant_impulse_response} presents the proposed phase-coherent CIR construction framework,
including delay and phase referencing based on a stable LoS-path reference.
Section~\ref{sec:Delay-Domain-Sensing-Experiments} demonstrates
bistatic range estimation through human gait and respiration experiments.
%Section~\ref{sec:Time-Domain-Preprocessing}
%addresses time-domain preprocessing methods for uniform resampling toward Doppler estimation.
Section~\ref{sec:delay-doppler-estimation} presents delay--Doppler response estimation results for moving targets.
%Section~\ref{sec:discussion} provides additional discussion,
%followed by
Section~\ref{sec:Conclusion} provides concluding remarks.

%11n, linux csi tool \cite{Halperin2011SIGCOMM} 以降

%\hrulefill

\section{Related Work}
\label{sec:Related-Work}

We review \mbox{Wi-Fi} sensing studies relevant to this work,
followed by reference-aided \mbox{Wi-Fi} radar approaches.
Estimation of propagation distance or time-of-flight based on statistical or
geometry-based features of CSI \cite{Wu2016UBICOMP,Tamai2025TVT}
is outside the scope of this review,
as such studies primarily aim at inferring static distance or position from
channel statistics, rather than resolving delay--Doppler responses
from phase-coherent CIRs.

\subsection{Doppler-Based Wi-Fi Sensing}
\label{ssec:Doppler-Based-Wi-Fi-Sensing}

This subsection focuses on \mbox{Wi-Fi} sensing studies that primarily perform
Doppler-related spectral analysis without explicit reference signals,
which are most closely related to the present work.
In these approaches, spectral analysis techniques such as the \gls{stft}
are typically applied to the magnitude of the CFR or its
low-dimensional representations.
This design choice is mainly due to the fact that the phase of the CFR
is generally unstable across packet acquisitions on COTS \mbox{Wi-Fi} devices,
making it difficult to reliably exploit phase information over time.
Although antenna-wise or subcarrier-wise phase differences have been
successfully exploited for localization
\cite{Kotaru2015SIGCOMM,Vasisht2016NSDI},
such phase-based quantities are not typically treated as time-domain
signals for subsequent time--frequency spectral analysis.

In \cite{Wang2016UbiCompRespiration}, respiration sensing is performed by
selecting subcarriers whose CFR magnitudes exhibit
target-induced periodic variations.
Spectral analysis is then applied to the selected magnitude-domain
time series to estimate the respiration rate.

WiFiU \cite{Wang2016UbiCompGait} adopts a \gls{pca}-based dimensionality
reduction of the CFR magnitude, followed by STFT on the %resulting
PC scores to extract Doppler-related information.
Widar \cite{Qian2017Mobihoc} further interprets the dominant spectral
components as path length change rates, and estimates target velocity and location through an explicit geometric model.
In both approaches, the Doppler-related
quantities are derived from magnitude-domain spectral analysis.

In particular, Li \emph{et al.} \cite{Li2021WCOM} provide an insightful
analysis closely related to the lack of stable phase reference addressed in this
work.
The authors contrast \mbox{Wi-Fi} sensing and \mbox{Wi-Fi} radar
by clarifying what physical quantities are estimated in each case.
They show that magnitude-based CFR processing inherently loses Doppler
sign information and physically meaningful phase evolution
because magnitude-only processing removes the phase evolution induced by
propagation path-length changes.
As a result, the extracted spectral components reflect periodic signal
strength fluctuations rather than a true Doppler frequency defined
by wave propagation.

%They show that magnitude-based CFR processing inherently loses Doppler
%sign information and physical phase interpretability because
%magnitude-only processing removes the phase evolution induced by
%propagation path-length changes.

\begin{table*}[t]
 \centering
 \caption{Positioning of This Work with Respect to Existing Studies Including ISAC-Enabling Capability}
 \begin{tabular}{cccccc}
  \toprule
  Study & Tx & Rx & Reference Signal (Fig.~\ref{fig:ref_schemes}) & ISAC-enabling & Standard \\
  \midrule
  Typical Wi-Fi sensing
  \cite{Wang2016UbiCompGait,Wang2016UbiCompRespiration,Qian2017Mobihoc}
    & COTS
    & COTS
    & $\times$
    & \checkmark
    & IEEE 802.11n \\

  \cite{Colone2012AES,Falcone2012AES,Pastina2015VT} 
    & COTS
    & Instrument Rx
    & (a) Wired reference 
    & $\times$
    & IEEE 802.11g \\

  \cite{Guo2008Radar,Chetty2012GRS}
    & COTS
    & Instrument Rx
    & (b) Reference antenna
    & $\times$
    & IEEE 802.11g \\

  \cite{Keerativoranan2018,Qiu2023IOT}         
    & COTS
    & COTS
    & (a) Wired reference
    & $\times$
    & IEEE 802.11n \\

  \cite{Song2020IOT}
    & COTS
    & COTS
    & (b) Reference antenna
    & $\times$
    & IEEE 802.11n \\

  \cite{Yamamoto2025EuCAP}
    & COTS
    & COTS
    & (b) Reference antenna
    & $\times$
    & IEEE 802.11ax \\

%  \cite{Yamamoto2025EuCAP}  
%    & COTS Wi-Fi device 
%    & CIR 
%    & Reference antenna \\

  \textbf{This work}
    & COTS
    & COTS
    & (c) \textbf{LoSRef: OTA LoS-path referencing}
    & \checkmark
    & \textbf{IEEE 802.11ax} \\
  \bottomrule
 \end{tabular}
 \label{tab:positioning}
\end{table*}

\subsection{Reference-Aided Wi-Fi Radar}
\label{ssec:wifi-radar}

Table~\ref{tab:positioning} summarizes the positioning of this work
with respect to representative \mbox{Wi-Fi} sensing and reference-aided \mbox{Wi-Fi} radar studies, highlighting differences in Tx and Rx types, 
reference signals, ISAC-enabling capability, and supported standards.
Here, ISAC-enabling indicates whether such functionality can be realized within typically deployed \mbox{Wi-Fi} systems without additional reference hardware.
In \cite{Guo2008Radar,Chetty2012GRS}, a bistatic radar architecture
is adopted, where a COTS \mbox{Wi-Fi} Tx and a measurement-grade Rx
are used.
The Rx employs separate reference and surveillance antennas,
as illustrated in Fig.~\ref{fig:ref_schemes}(b),
with the antenna configuration such that target-induced variations
are captured in the Tx--surveillance path.
The target-induced Doppler frequency is then extracted by correlating the reference and
surveillance signals.

As an alternative reference acquisition strategy,
\cite{Colone2012AES,Falcone2012AES,Pastina2015VT} inserted a directional
coupler between a COTS \mbox{Wi-Fi} Tx and its antenna to obtain a
wired reference signal, as illustrated in Fig.~\ref{fig:ref_schemes}(a).
It should be noted that in all these studies \cite{Guo2008Radar,Chetty2012GRS,Colone2012AES,Falcone2012AES,Pastina2015VT}, the Rx is instrument-grade, and sensing is performed directly on received signals rather than on CSI measurements obtained from COTS \mbox{Wi-Fi} devices.
Moreover, these studies largely predate the widespread availability of CSI on COTS \mbox{Wi-Fi} hardware, which emerged later with IEEE 802.11n and CSI extraction tools \cite{Halperin2011SIGCOMM}.
%, followed by subsequent standards.
In this sense, \mbox{Wi-Fi} is exploited merely as a signal of opportunity, without leveraging the CSI natively observed by COTS \mbox{Wi-Fi} devices.
Nevertheless, these works demonstrate that \mbox{Wi-Fi} signals can fundamentally support radar functionality.

Building on these reference-based architectures, several
studies have demonstrated fully COTS-based implementations,
in which both the Tx and Rx are COTS \mbox{Wi-Fi} devices,
while still exploiting additional reference hardware for phase calibration.
In \cite{Keerativoranan2018,Qiu2023IOT}, a wired reference is employed.
This enables explicit Doppler frequency estimation for fine-grained hand gesture
recognition \cite{Keerativoranan2018} and respiration monitoring in dynamic
environments \cite{Qiu2023IOT}.

Similarly, WiEps \cite{Song2020IOT} and SoundiFi \cite{Yamamoto2025EuCAP} adopt a reference antenna
to achieve phase-stable CSI measurements under a fully COTS-based setup.
These capabilities are exploited for dielectric property measurement of materials
\cite{Song2020IOT} and for channel sounding \cite{Yamamoto2025EuCAP}, respectively.
These results indicate that wired or antenna-based reference signals
remain an effective and reliable means of phase-coherent radar-like analysis
using \mbox{Wi-Fi} devices.
Note that although not all of these studies explicitly use the term ``radar,''
their architectures are consistent with bistatic radar principles,
as they rely on explicit reference signals to enable phase-coherent analysis.

In contrast to existing reference-aided \mbox{Wi-Fi} radar systems,
which rely on wired references or dedicated reference antennas,
the LoSRef exploits the OTA Tx--Rx LoS path.
As a result, phase-coherent bistatic radar operation is achieved
using only a single-antenna, single-port Rx,
without requiring multi-port Rx architectures,
thereby enabling integration into typical
\mbox{Wi-Fi} systems and facilitating ISAC operation.

\section{CIR Representation and Reference Ambiguity Due to Free-Running Clocks}
\label{sec:CIR-Representation}

This section discusses the reference ambiguity in practical \mbox{Wi-Fi} systems,
which fundamentally prevents phase-coherent analysis and
physically meaningful time-domain processing.

\subsection{CIR Representation and Doubly Selective Channels}
\label{ssec:CIR-Model}

The equivalent low-pass representation of the CIR can be described using a \gls{tdl} model \cite{Molisch2023}.
Under this model, the CIR at measurement time $t$ is expressed as a superposition of discrete propagation paths, each characterized by a complex path gain and a propagation delay, as
\begin{align}
 h(\tau,t) = \sum_{\ell=1}^{L} \eta_\ell \, \mathrm{e}^{-\mathrm{j}2\pi f_{\mathrm{c}} \tau_\ell} \delta(\tau - \tau_\ell),
 \label{eq:TDL}
\end{align}
where \( L \) denotes the number of paths, $\eta_\ell$ and $\tau_\ell$ represent the complex amplitude and delay of the $\ell$-th path, respectively, and $f_{\mathrm{c}}$ is the carrier frequency.
Here, $\tau$ denotes the signal propagation delay measured from the
transmission time,
and $t$ denotes the observation time of the CIR, corresponding to successive channel measurements rather than a continuous-time variable.
This representation captures the multipath structure of the wireless channel in the delay domain, where the phase of each path reflects the propagation path length.

In dynamic environments, the underlying wireless channel is
\emph{doubly selective}, exhibiting variations across both delay and time.
The CIR $h(\tau,t)$ captures such joint delay--time variations.

\subsection{Reference Ambiguity Due to Free-Running Clocks}
\label{ssec:Reference-Ambiguity}

In \mbox{Wi-Fi} systems, where the Tx and Rx employ independent, free-running clocks, neither a common timing reference nor an absolute phase reference is available between them.
As a result, the CIR observed at the Rx does not preserve the absolute delay
reference to the transmission timing, nor does it retain a phase-coherent representation across successive observations.
Accordingly, the observed CIR $h_\mathrm{raw}(\tau,t)$ is related to the
physical CIR $h(\tau,t)$ through an unknown delay offset and an
unknown phase offset, and is given by
\begin{align}
 h_\mathrm{raw}(\tau,t) = h(\tau + \alpha_t,t)\, \mathrm{e}^{\mathrm{j}\beta_t},
 \label{eq:h_and_hraw}
\end{align}
where $\alpha_t$ and $\beta_t$ are the delay and phase offsets,
respectively, which may vary across observation times $t$ \cite{Kotaru2015SIGCOMM,Vasisht2016NSDI}.
These offsets originate from free-running clocks at the Tx and Rx,
and prevent direct use of absolute delay and phase information,
thereby precluding phase-coherent analysis of the doubly selective channel in practical \mbox{Wi-Fi} systems without a shared reference.
Note that the delay offset $\alpha_t$ of the devices used in this study
has been experimentally characterized in \cite{Yamamoto2025EuCAP}.

When \eqref{eq:h_and_hraw} is viewed in the frequency domain,
the resulting complex-valued CFR $H_{\mathrm{raw}}(f,t)$,
where $f$ denotes the OFDM subcarrier frequency,
is inherently discontinuous with respect to time $t$ due to these unknown offsets, which precludes
direct temporal or spectral analysis.
To circumvent this issue, many \mbox{Wi-Fi} sensing studies,
including \cite{Wang2016UbiCompRespiration},
operate on the magnitude of the CFR, $|H_{\mathrm{raw}}(f,t)|$.
Taking the magnitude removes these unknown offsets,
yielding a smooth time series that is amenable
to spectral analysis.
This convenience, however, comes at the cost of discarding the
phase evolution associated with wave propagation,
thereby precluding any phase-coherent interpretation of
propagation delay or Doppler frequency.

In contrast, in bistatic radar systems where a reference signal is available,
a phase-coherent CIR \(h(\tau,t)\) can be recovered.
This temporal continuity enables the use of the Fourier-transform
duality between $h(\tau,t)$ and physically interpretable
delay--Doppler representations \cite{Bello1963,Molisch2023}.

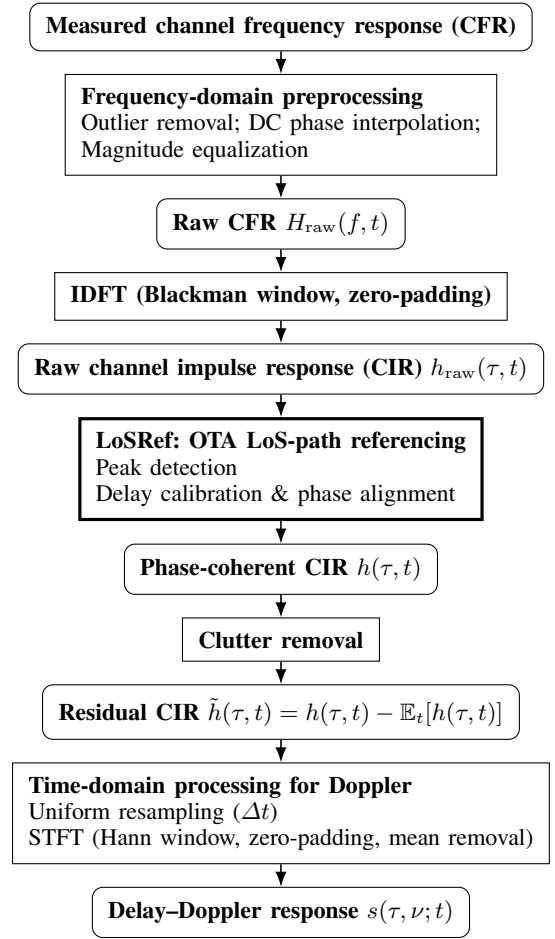
\begin{figure}[t]
\centering
\begin{tikzpicture}[
  font=\small,
  node distance=3mm and 10mm,
  % ===== Styles =====
  data/.style={
    draw,
    rounded corners=4pt,
    align=left,
    inner xsep=6pt,
    inner ysep=5pt
  },
  proc/.style={
    draw,
    rectangle,
    align=left,
    inner xsep=6pt,
    inner ysep=5pt
  },
  key/.style={ % key proposed block (shape only: thicker border)
    proc,
    very thick
  },
  arrow/.style={-Latex, line width=0.6pt}
]

% ===== Nodes =====
\node[data] (cfr) {\textbf{Measured channel frequency response (CFR)}};

\node[proc, below=of cfr] (fdprep) {\textbf{Frequency-domain preprocessing}\\% (\S\ref{ssec:Frequency-Domain-Preprocessing})\\
Outlier removal; DC phase interpolation;\\
Magnitude equalization};
%\(\bullet\) Outlier removal\\
%\(\bullet\) DC phase interpolation\\
%\(\bullet\) Spectral-edge correction\\
%\(\bullet\) Magnitude equalization};

\node[data, below=of fdprep] (cfrraw) {\textbf{Raw CFR} $H_\mathrm{raw}(f,t)$};
 
\node[proc, below=of cfrraw] (idft) {\textbf{IDFT (Blackman window, zero-padding)}};
% (\S\ref{ssec:CIR_Construction})\\
%(delay-domain oversampling)};

\node[data, below=of idft] (cirraw) {\textbf{Raw channel impulse response (CIR)} $h_{\mathrm{raw}}(\tau,t)$};

\node[key, below=of cirraw] (ota) {\textbf{LoSRef: OTA LoS-path referencing}\\% (\S\ref{ssec:Delay-Calibration})\\
Peak detection\\
Delay calibration \& phase alignment};

\node[data, below=of ota] (circal) {\textbf{Phase-coherent CIR} $h(\tau,t)$};

\node[proc, below=of circal] (scr) {\textbf{Clutter removal}}; % (\S\ref{ssec:Static-Component-Removal})\\
%\(\overline{h}(\tau)=\mathbb{E}_t[h(\tau,t)]\),

\node[data, below=of scr] (rcir) {
\textbf{Residual CIR} \(\tilde{h}(\tau,t)=h(\tau,t)-\mathbb{E}_t[h(\tau,t)]\)};

\node[proc, below=of rcir] (tdproc) {\textbf{Time-domain processing for Doppler}\\
Uniform resampling (\(\varDelta t\))\\
STFT (Hann window, zero-padding, mean removal)};
 
%\node[proc, below=of scr] (grid) {\textbf{Uniform time resampling}\\% (\S\ref{ssec:Uniform-Resampling})\\
%Median grid spacing \(\varDelta t\)\\
%Nearest-neighbor assignment (duplication allowed)};
%
%\node[proc, below=of grid] (stft) {\textbf{STFT along time (per delay bin)}\\% (\S\ref{ssec:Range--Doppler-Estimation})\\
%Hann window; zero-padding; mean removal};
%% (Doppler-domain oversampling)\\
%%Mean removal per STFT segment};

\node[data, below=of tdproc] (dd) {\textbf{Delay--Doppler response} $s(\tau,\nu;t)$};

% ===== Arrows =====
\draw[arrow] (cfr) -- (fdprep);
\draw[arrow] (fdprep) -- (cfrraw);
\draw[arrow] (cfrraw) -- (idft);
\draw[arrow] (idft) -- (cirraw);
\draw[arrow] (cirraw) -- (ota);
\draw[arrow] (ota) -- (circal);
\draw[arrow] (circal) -- (scr);
\draw[arrow] (scr) -- (rcir);
\draw[arrow] (rcir) -- (tdproc);
\draw[arrow] (tdproc) -- (dd);
%\draw[arrow] (grid) -- (stft);
%\draw[arrow] (stft) -- (dd);

\end{tikzpicture}
\caption{Processing flow from measured CFR to the delay--Doppler response.
The LoSRef exploits the Tx--Rx LoS path as a delay and phase reference, enabling phase-coherent alignment of CIR across acquisitions.}
\label{fig:pipeline}
\end{figure}

\section{Phase-Coherent Time-Variant Impulse Response Estimation}
\label{sec:time-variant_impulse_response}

The offsets $\alpha_t$ and $\beta_t$ in \eqref{eq:h_and_hraw}
are unknown and vary across observation times, which prevents
phase-coherent alignment of CIRs across successive packet acquisitions.
In contrast, the delay of the LoS-path component in \eqref{eq:TDL}
is determined by the Tx--Rx distance and is therefore expected to
remain constant as long as the Tx and Rx are stationary.
Exploiting this temporal stability, the LoSRef uses the
LoS-path component as an \gls{ota} reference to mitigate the effects
of these unknown offsets and to enable
phase-coherent time-variant CIR estimation.
Fig.~\ref{fig:pipeline} summarizes the processing flow of the proposed framework,
incorporating LoSRef,
from the measured CFR to the delay--Doppler response,
and each processing block is detailed in the following subsections.

\subsection{CIR Construction for Robust LoS-Path Identification}
\label{ssec:CIR_Construction}

Recall that \( H_{\mathrm{raw}}(f,t) \) denotes the CFR observed at time \( t \).
The corresponding CIR is obtained by transforming
the CFR from the frequency domain to the delay domain via an inverse Fourier
transform, given by
\begin{align}
 h_{\mathrm{raw}}(\tau,t)
 = \mathcal{F}^{-1} [ H_{\mathrm{raw}}(f,t) ].
 \label{eq:H2h}
\end{align}
In practice, the CFR is further processed by applying windowing and effective
bandwidth extension, as described below.
Strictly speaking, the term \( H_{\mathrm{raw}}(f,t) \) in
(\ref{eq:H2h}) should therefore be interpreted as the preprocessed CFR.

%It should be noted that
The resulting CIR includes the phase and gain of the transceivers and the radio channel,
and is convolved with the impulse response of the applied frequency-domain window.
For simplicity, this windowed version is hereafter referred to as the CIR.
Similarly, although the observations are discrete and
the transformation should be written as an \gls{idft}, we retain the simplified representation in (\ref{eq:H2h}) to
avoid notational complexity.

To construct a stable delay-domain CIR, frequency-domain windowing and
delay-domain oversampling are applied, following the same principle as in
our prior \mbox{Wi-Fi} channel sounding work \cite{Yamamoto2025EuCAP},
with window and oversampling parameters adapted to the present setup.
Specifically, a Blackman window is used prior to the inverse Fourier transform
to suppress delay-domain sidelobes.
In addition, delay-domain oversampling with a factor of $\kappa = 32$ is
performed via zero padding in the frequency domain. 
It is emphasized that such oversampling does not improve the intrinsic
delay resolution, which remains limited by the signal
bandwidth ($1/B = \SI{6.3}{\nano\second}$), but is nevertheless
essential for achieving accurate and temporally stable peak detection
under discretized delay sampling.

%\cite[Section III-C]{Yamamoto2025EuCAP},

\subsection{LoSRef-Based Delay Calibration and Phase Alignment}
\label{ssec:Delay-Calibration}

Let the known LoS \gls{ota} Tx--Rx propagation distance \( d_{\text{Tx--Rx}} \) be used as the reference distance \( d_{\mathrm{ref}} \).
For clarity, this OTA LoS-path referencing scheme is hereafter simply referred to as LoSRef.
While the measurement accuracy of \( d_{\mathrm{ref}} \) may warrant careful
discussion depending on the target application, this issue is beyond the scope
of this work, since \( d_{\mathrm{ref}} \) is used solely for absolute delay-axis
alignment.
Doppler frequency estimation, which is based on the temporal derivative of the
propagation phase, is therefore independent of the absolute value of
\( d_{\mathrm{ref}} \), as long as it is constant over time.
The propagation delay of the LoS path relative to the transmitted signal is
given by \( \tau_{\mathrm{ref}} = d_{\mathrm{ref}}/c \),
where \( c \) denotes the speed of light.
In an ideal delay-synchronized system, the LoS-path component would therefore
appear at \( \tau_{\mathrm{ref}} \) in the CIR.

In practice, however, this alignment is not observed, because the Tx and Rx
are not delay-synchronized and the delay axis of the measured CIR is not
referenced to the actual packet transmission timing.
As a result, the absolute delay origin varies across packet acquisitions.
To address this issue, the delay axis of the CIR is calibrated by identifying
the dominant component, assumed to correspond to the LoS path, and aligning
its delay with the reference delay \( \tau_{\mathrm{ref}} \).
This procedure establishes a common delay reference across packet acquisitions.%
\footnote{%
Strictly speaking, the identified peak corresponds to a delay bin dominated by
the LoS path rather than an isolated propagation path.
}
Specifically, the delay calibration is
performed to satisfy
\begin{align}
\argmax_{\tau} \, \lvert h(\tau,t) \rvert = \tau_{\mathrm{ref}}.
\end{align}

\begin{figure}[t]
 \includegraphics[width=\linewidth]{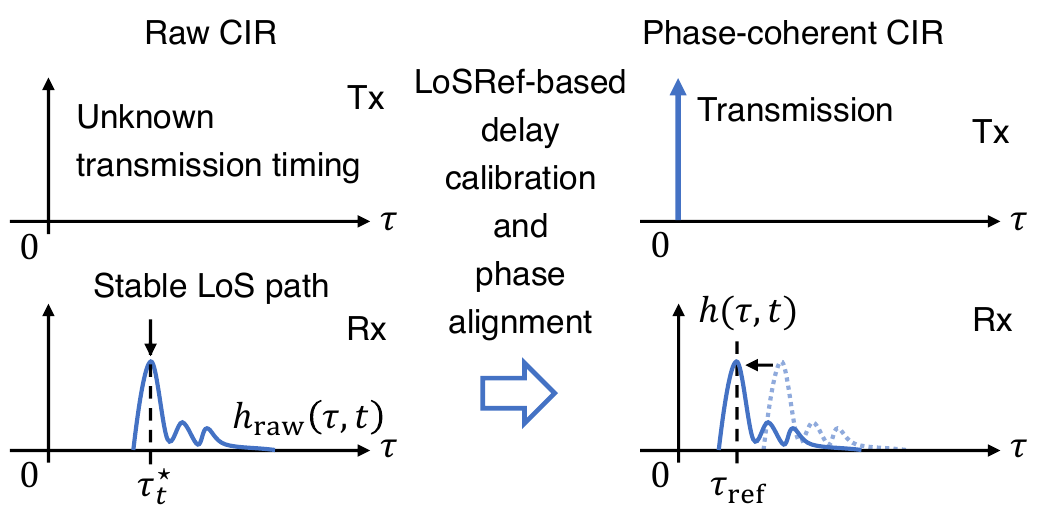}
 \caption{LoSRef: OTA LoS-path referencing for delay calibration and phase alignment.
 The dominant LoS-path component is used as an OTA reference to compensate unknown offsets arising from free-running clocks.}
 \label{fig:250320_key_idea}
\end{figure}

This calibration is carried out as follows.
For each time index \( t \), the delay and phase of the strongest signal in
the raw CIR \( h_{\mathrm{raw}}(\tau,t) \) are first identified as
\begin{align}
 \tau_t^\star &= \argmax_{\tau} \, \lvert h_{\mathrm{raw}}(\tau, t) \rvert, \\
 \theta_t^\star &= \arg \, h_{\mathrm{raw}}(\tau_t^\star, t).
\end{align}
The calibrated CIR is then obtained by compensating for both the delay offset
and the phase rotation of this strongest signal,
 which jointly restores phase coherence across successive CIR measurements,
given by
\begin{align}
 h(\tau,t)
 = h_{\mathrm{raw}} (\tau - (\tau_{\mathrm{ref}} - \tau_t^\star), t )
 \, \ee^{-\jj \theta_t^\star},
\end{align}
as illustrated in Fig.~\ref{fig:250320_key_idea}.
The above calibration procedure follows the same mathematical formulation as
our prior channel sounding work \cite{Yamamoto2025EuCAP}.
In addition, the magnitude of the CIR is normalized such that
\( \lvert h(\tau_{\mathrm{ref}}, t) \rvert \) matches the free-space path gain
corresponding to the reference distance \( d_{\mathrm{ref}} \).
The resulting CIR \( h(\tau,t) \) represents the complex
path gain associated with each propagation path.
In this calibration procedure, the delay shift by
\( \tau_{\mathrm{ref}} - \tau_t^\star \) is referred to as
\textit{delay calibration}, while the phase rotation by
\( \ee^{-\jj \theta_t^\star} \) is referred to as
\textit{phase alignment}.
Through phase alignment, the phase of each multipath component
is expressed relative to that of the dominant (LoS-path) component.
As a result, the phase of the CIR becomes continuous across successive
acquisitions, enabling phase-coherent time-variant and Doppler analysis.

The LoSRef-based calibration relies on the assumption that the LoS-path
component remains dominant and is well resolvable in the CIR.
Scenarios in which this assumption is violated,
such as extremely close target proximity to the Tx--Rx pair, are beyond the
scope of this work.

\subsection{Clutter Removal via LoSRef-based CIR Alignment}
\label{ssec:Static-Component-Removal}

As in \eqref{eq:TDL}, the observed CIR \( h(\tau,t) \) can be modeled as a superposition of
static multipath components, hereafter referred to as clutter, 
and time-varying components induced by target motion.
%The observed CIR \( h(\tau,t) \) can be modeled, as in \eqref{eq:TDL}, as a superposition of
%static multipath components, hereafter referred to as clutter, and time-varying components induced by target motion.
In indoor environments, the CIR is typically dominated by strong clutter, which tends to mask relatively weak motion-induced
variations associated with human targets.

To isolate the time-varying component, clutter is removed using a temporal averaging approach.
Specifically, the clutter is estimated by taking the temporal average of the CIR over the observation interval, i.e.,
\begin{align}
\overline{h}(\tau) \coloneqq \mathbb{E}_t [ h(\tau,t) ].
\end{align}
The time-varying component is then extracted by subtracting the estimated clutter from the original CIR:
\begin{align}
 \tilde{h}(\tau,t) \coloneqq h(\tau,t) - \overline{h}(\tau).
\end{align}
The resulting signal, referred to as the \textit{residual CIR}
\( \tilde{h}(\tau,t) \), is used for subsequent sensing and analysis.

In radar signal processing, clutter removal is a standard operation in which
static components, including the direct path, are suppressed.
In contrast, the proposed LoSRef-based framework differs in that the
LoS-path component is first exploited as a delay and phase
reference to achieve CIR alignment, and subsequently suppressed
through clutter removal.

Note that, in conventional \mbox{Wi-Fi} sensing systems, direct processing of
complex-valued CIRs is not physically meaningful due to the lack of a shared delay and phase
reference between the Tx and Rx,
%due to the lack of delay synchronization between the Tx and Rx,
which prevents meaningful temporal averaging.
For example, Widar \cite{Qian2017Mobihoc} extracts motion-related components
by applying \gls{pca} to the CFR magnitude, followed by time--frequency
spectral analysis.
By contrast, the proposed LoSRef-based approach aligns CIRs on a common
delay axis using a stable LoS-path reference, enabling phase-coherent
temporal averaging and delay-domain clutter removal, which may
be advantageous for slowly varying motions.

\section{Experimental Analysis of OTA Phase-Referenced Channel Impulse Responses}

%\section{Experimental Analysis of Phase-Coherent Delay-Domain Sensing}
%\section{Delay-Domain Sensing Experiments}
\label{sec:Delay-Domain-Sensing-Experiments}

To validate the proposed framework described in
Section~\ref{sec:time-variant_impulse_response}, we conducted two
representative experimental scenarios designed to capture both (i) coarse delay-domain
magnitude variations induced by translational human gait and (ii)
sub-wavelength, phase-sensitive variations induced by respiration.

\subsection{Experimental Equipment}

\begin{figure}[t]
 \centering
 \includegraphics[width=.45\linewidth]{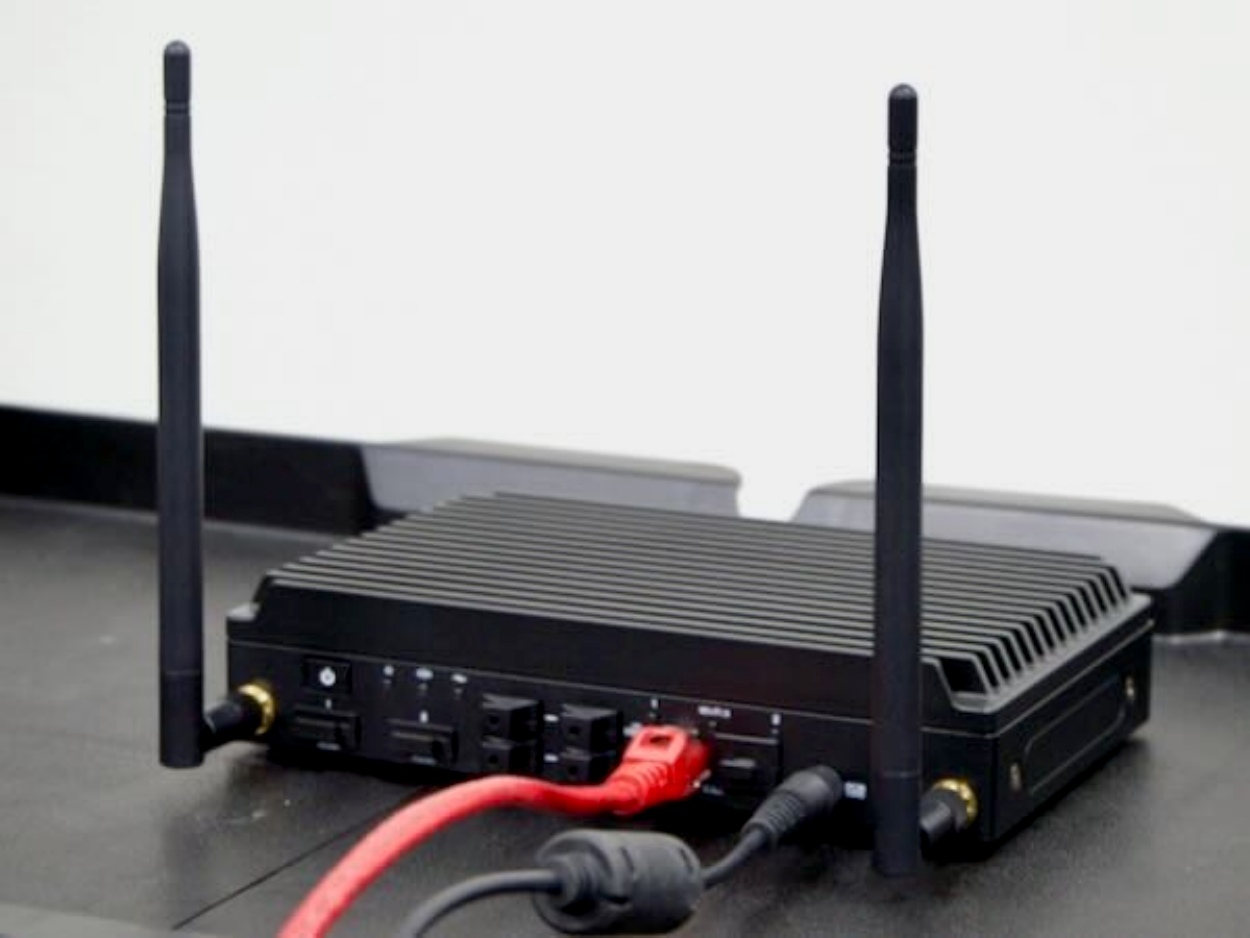}
 \caption{A COTS Wi-Fi device used as the Tx or Rx.}
 \label{fig:NUC}
\end{figure}

\begin{table}[t]
 \centering
 \caption{Transmission and CIR Construction Parameters}
 \begin{tabular}{cc}
  \toprule
  Standard & IEEE 802.11ax \cite{ieee80211ax} \\
  Number of Tx/Rx antennas (used) & 1 / 1 \\
  %MIMO configuration & SISO \\
  Center frequency $f_\mathrm{c}$ & \SI{5570}{\mega\hertz} \\%5570\,MHz \\
  Wavelength $\lambda$ & \SI{53.8}{\milli\meter} \\
  Bandwidth $B$ & \SI{160}{\mega\hertz} \\ %160\,MHz \\
  Subcarrier spacing $\varDelta f$ & \SI{78.125}{\kilo\hertz} \\ %78.125\,kHz \\
  Number of observed subcarriers & \num{1992} \\
  Number of subcarriers after interpolation & \num{2025} \\
  Subcarrier indices & \num{-1012}, \num{-1011}, \ldots, \num{1012} \\
  Nominal packet transmission interval $\varDelta t$ & \SI{1}{\milli\second} \\
  Number of FFT points w/o interpolation & \num{2048} \\
  Delay-domain interpolation rate $\kappa$ & \num{32} \\
  Frequency-domain window & Blackman \\
  %Interpolation rate in Doppler domain & \num{32} \\
  %Coaxial cables & 5\,m \\
  \bottomrule
 \end{tabular}
 \label{tab:parameters}
\end{table}

As the Tx and Rx, we used two ASUS NUC 13 Rugged-Tall devices, shown in Fig.~\ref{fig:NUC}, each equipped with an Intel AX210 \mbox{Wi-Fi} card and running PicoScenes \cite{Jiang2022IOT} for CSI acquisition.
Only a single external antenna was used at each device, resulting in a single-input single-output channel.
In the following, the Tx and Rx positions are defined by the locations of the corresponding antennas.
The devices were connected to a control PC via Ethernet and remotely operated to coordinate packet transmission and reception.
The Tx transmitted packets using the parameters summarized in Table~\ref{tab:parameters}.
The Rx received these packets and acquired the corresponding CSI, along with the RSSI.
The nominal packet transmission interval was set to \SI{1}{\milli\second}; however, the actual inter-packet intervals observed at the Rx are not perfectly uniform.
The characteristics of this timing irregularity are discussed in Section~\ref{sec:delay-doppler-estimation}.

For ground-truth position reference in the human gait experiment, HTC VIVE Tracker 3.0 \cite{merker2023sensors} based on infrared optical tracking, together with SteamVR Base Station 2.0, were attached to the moving human target's head and to the Tx and Rx.

\subsection{Frequency-Domain Preprocessing}
\label{ssec:Frequency-Domain-Preprocessing}

In addition to the preprocessing provided by the PicoScenes Python Toolbox \cite{Jiang2022IOT}, such as interpolation of unobserved pilot subcarriers, further frequency-domain preprocessing steps were applied, partially following the procedure described in Section VI-C of \cite{Yamamoto2025EuCAP}, with additional preprocessing steps introduced in this study.
After applying these preprocessing steps, the resulting frequency-domain channel representation is denoted as \( H_{\mathrm{raw}}(f,t) \).
The details of the preprocessing procedure are as follows.

As an outlier removal procedure, CFR samples that exhibit abrupt and
extremely low RSSI are treated as outliers and excluded
from further analysis.
In the considered bistatic configuration, target motion affects the CFR,
but is not expected to induce large changes in the overall received power,
which is dominated by the LoS path.
Accordingly, sudden power drops exceeding \SI{10}{\decibel}
are regarded as outliers
since the received RSSI, reported with \SI{1}{\decibel} granularity,
typically varies only on the order of a few decibels.

\begin{figure}[t]
 \subfloat[][The inter-subcarrier phase difference is derived from the unwrapped phase of the CFR, $H(f,t)$.]{%
 \includegraphics[width=\linewidth]{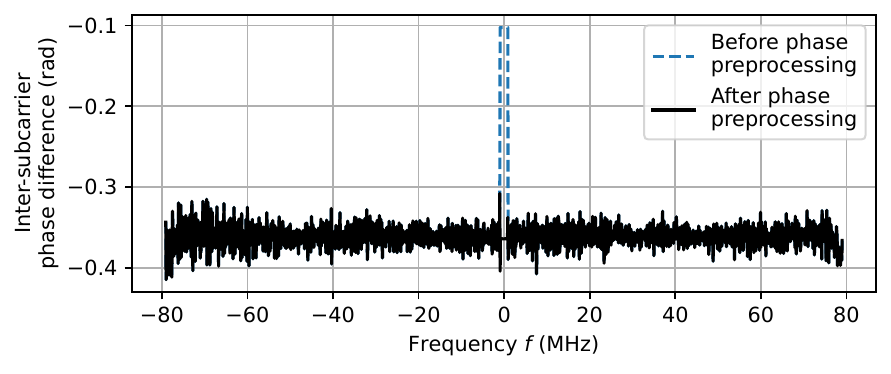}%
 \label{fig:cfr_phase}
 }

 \subfloat[][The magnitude is normalized by the RMS value after spectral-edge power normalization and the \(\pm\SI{60}{\mega\hertz}\) attenuation correction.]{%
 \includegraphics[width=\linewidth]{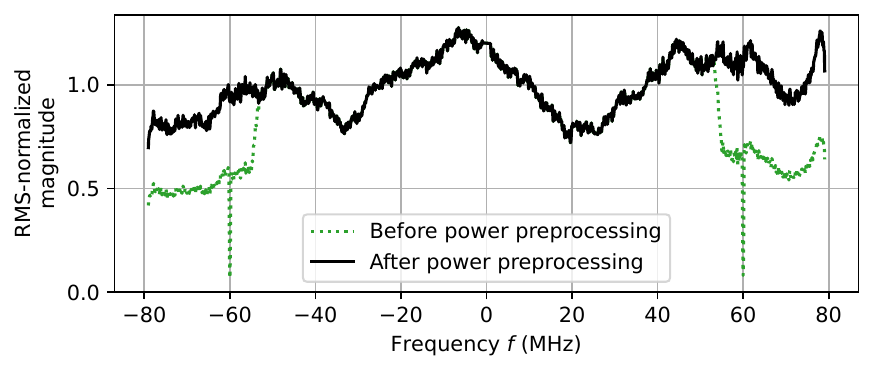}%
 \label{fig:cfr_magnitude}
 }

 \caption{Impact of phase and power preprocessing on the CFR.}
 \label{fig:preprocessing}
\end{figure}

Phase interpolation \cite{Yamamoto2025EuCAP} was applied to the DC subcarriers (indices \num{-11} to \num{11}), which carry no transmitted signal in IEEE 802.11ax \cite{ieee80211ax}.
As shown in Fig.~\ref{fig:preprocessing}\subref{fig:cfr_phase}, prior to this correction, ignoring the inherent $2\pi$ phase ambiguity results in anomalous inter-subcarrier phase differences at the DC subcarriers.
The inter-subcarrier phase difference is defined as
\begin{align}
\arg \parens*{
 \frac{H_\mathrm{raw}((k+1)\varDelta f,t)}{H_\mathrm{raw}(k\varDelta f,t)}
 },
\end{align}
where $k \in \{ \num{-1012}, \num{-1011}, \ldots, \num{1012} \}$ denotes the subcarrier index, with the corresponding frequency given
by $f = k\varDelta f$ and the subcarrier spacing
$\varDelta f = \SI{78.125}{\kilo\hertz}$.
To resolve this issue, the phase interpolation estimates the phase trend from neighboring subcarriers using the unwrapped phase.
The DC subcarrier phases are then reconstructed accordingly, as confirmed by the resulting phase differences in the DC region being consistent with those of the surrounding subcarriers.

As shown in Fig.~\ref{fig:preprocessing}\subref{fig:cfr_magnitude},
pronounced attenuation appears around frequency offsets of approximately \(\pm\SI{60}{\mega\hertz}\).
Following the same treatment as in \cite{Yamamoto2025EuCAP}, this attenuation is considered to originate from device-specific characteristics rather than wave propagation effects.
Accordingly, the CFR values at the corresponding subcarriers are excluded, and their magnitudes are interpolated from those of the neighboring subcarriers.

\begin{figure}[t]
 \includegraphics[width=\linewidth]{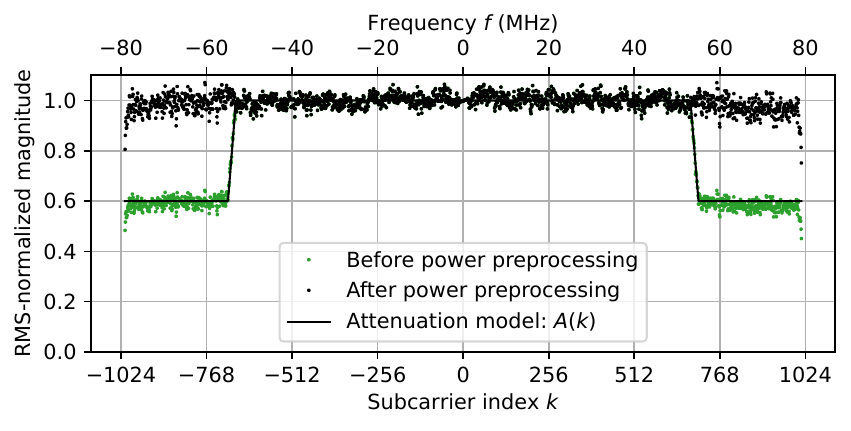}
 % 241204_CFRの補正
 \caption{Measured and compensated CFR magnitudes for a single snapshot with direct coaxial connection, with the estimated attenuation model overlaid.}
 \label{fig:correct_high_freq_magnitude}
\end{figure}

In addition, as observed in Fig.~\ref{fig:preprocessing}\subref{fig:cfr_magnitude},
the CFR magnitude exhibits noticeable attenuation for frequencies with $|f| > \SI{53}{\mega\hertz}$.
To quantitatively characterize this attenuation, the Tx and Rx were directly connected using coaxial cables and an attenuator, thereby forming a non-frequency-selective channel.
Fig.~\ref{fig:correct_high_freq_magnitude} shows an example of the measured CFR, where the magnitude is normalized by the RMS value over subcarriers with $|k| \leq \num{680}$.
The measured CFR magnitude is approximately unity for subcarriers with \(|k| \leq \num{680}\), while it decreases to about \num{0.6} for \(|k| \geq \num{704}\), with a nearly linear transition observed between these regions.
To capture this attenuation behavior, we introduce a piecewise attenuation model $A(k)$ in this study as
\begin{align}
A(k) =
\begin{cases}
1, & |k| \leq \num{680}, \\
-(|k|-\num{680})/60 + 1, & \num{680} < |k| < \num{704}, \\
\num{0.6}, & \num{704} \leq |k|,
\end{cases}
\end{align}
which approximates the observed magnitude trend and is overlaid in Fig.~\ref{fig:correct_high_freq_magnitude}.
Based on this model, the non-uniform frequency response is compensated by scaling the magnitude of each subcarrier by the inverse of $A(k)$.
This operation effectively equalizes the spectral magnitude, resulting in an approximately uniform CFR across all subcarriers, as shown in Fig.~\ref{fig:correct_high_freq_magnitude}.

We confirmed consistent frequency-domain attenuation and phase behavior
across three or more ASUS NUC devices.
Notably, different device sets were used in
Sections~\ref{ssec:range-estimation} and~\ref{ssec:Respiration},
yet comparable characteristics were observed.
This suggests that device-to-device variability is limited for the hardware
used in this study, and that the proposed preprocessing is practically
applicable under similar experimental conditions.

\subsection{Impact of Preprocessing on CIR Construction}

\begin{figure}[t]
 \includegraphics[width=\linewidth]{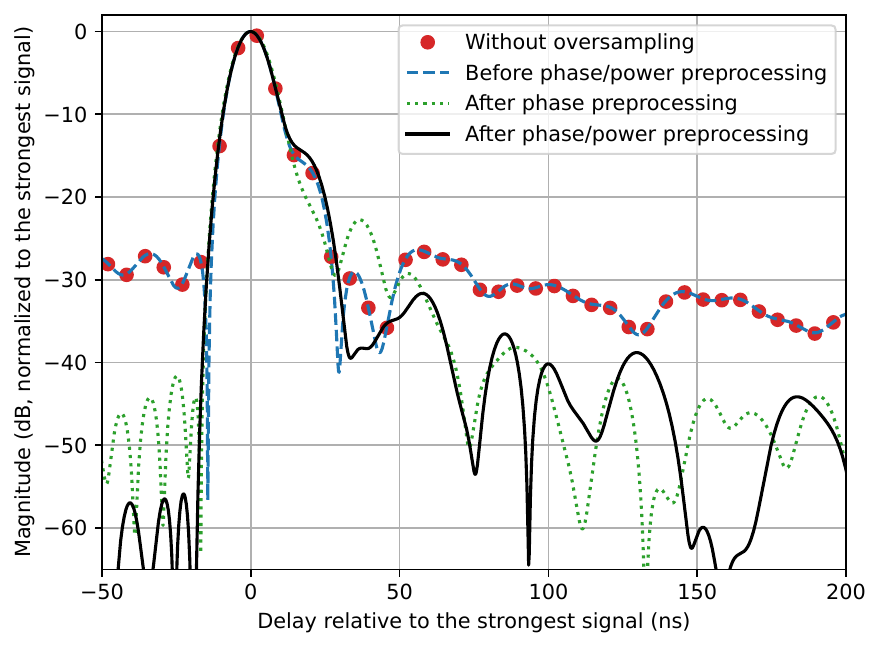}
 \caption{Impact of phase and power preprocessing on the CIR.
 %Windowing described in Section~\ref{ssec:CIR_Construction} is applied to all curves,
%whereas
 %Oversampling is applied to all curves except for the ``Without oversampling'' case.
 Note that the horizontal axis of the ``Without oversampling'' curve is manually shifted to match that of the ``Before phase/power preprocessing'' curve for visualization only.}
 \label{fig:impact_of_preprocessing}
\end{figure}

Fig.~\ref{fig:impact_of_preprocessing} examines the impact of preprocessing on the constructed CIR.
The delay axis is referenced to the strongest path, which is assumed to correspond
to the LoS path.
In the LoS Tx--Rx experimental environment considered here, negative delays
correspond to non-physical artifacts.
The figure also illustrates how delay-domain oversampling helps alleviate
discretization effects in peak delay estimation.

%Without oversampling, the CIR is sampled coarsely along the delay axis, causing
%the discrete samples to miss the true peak location.
%As this sampling offset varies across time, the apparent peak delay may fluctuate
%even for a static propagation path, hindering stable delay calibration.
%Unless otherwise stated, the following discussion refers to the CIR obtained
%after applying delay-domain oversampling.

Before phase/power preprocessing, a noticeable pre-LoS component was observed at approximately
\(-\num{26}\,\mathrm{dB}\), exceeding the expected Blackman-window sidelobe level
of about \(-\num{58}\,\mathrm{dB}\).
After preprocessing, such components were suppressed to the sidelobe level,
indicating effective mitigation of spurious responses.
Minor residual discrepancies may still remain due to non-ideal hardware effects.
%of about \(-\num{31}\,\mathrm{dB}\). % han

The wide bandwidth of \SI{160}{\mega\hertz} available in IEEE 802.11ax
yields a well-localized impulse response with suppressed sidelobes,
providing a stable and temporally consistent LoS-path peak.
From an estimation-theoretic perspective, a wider bandwidth also tightens the
Cram\'er--Rao lower bound for delay estimation \cite{Jin1995CRLB}, thereby improving
the achievable delay estimation accuracy.
This contrasts with many existing \mbox{Wi-Fi} sensing studies based on
\SI{20}{\mega\hertz} or \SI{40}{\mega\hertz} channels of IEEE 802.11n devices,
where limited bandwidth results in a less distinct delay-domain structure.
Consequently, the LoS-path component serves as a reliable physical reference
for LoSRef and subsequent phase-coherent delay-domain analysis.

\subsection{Range Estimation of Translational Human Gait}
\label{ssec:range-estimation}

\subsubsection{Experimental Setup and Range--Time Representation}

\begin{figure}[t]
 \centering
 \subfloat[][Geometry and back-and-forth straight-line gait trajectory obtained from the motion-tracking system.]{
 %Geometry and walking trajectory obtained from the motion-tracking system.]{
 \centering
 \includegraphics[width=.55\linewidth]{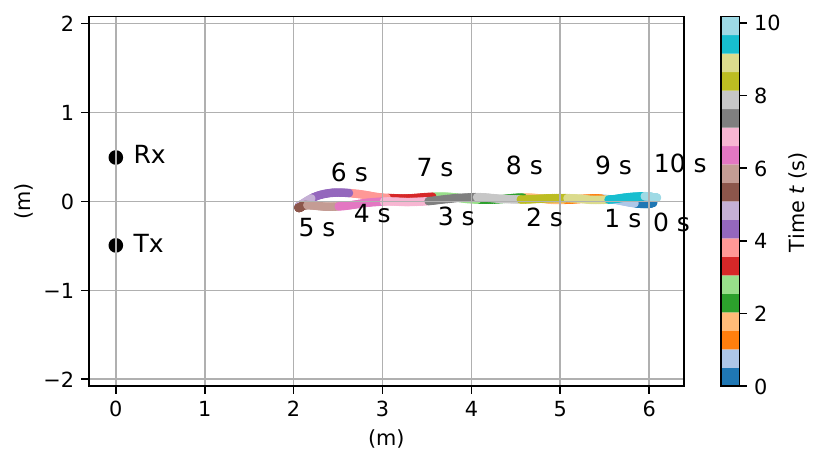}
 \label{fig:geometry}
 }
 \hfill
 \subfloat[][Experimental setup with the Tx--Rx pair and the human target.]{
 \centering
 \includegraphics[width=.35\linewidth]{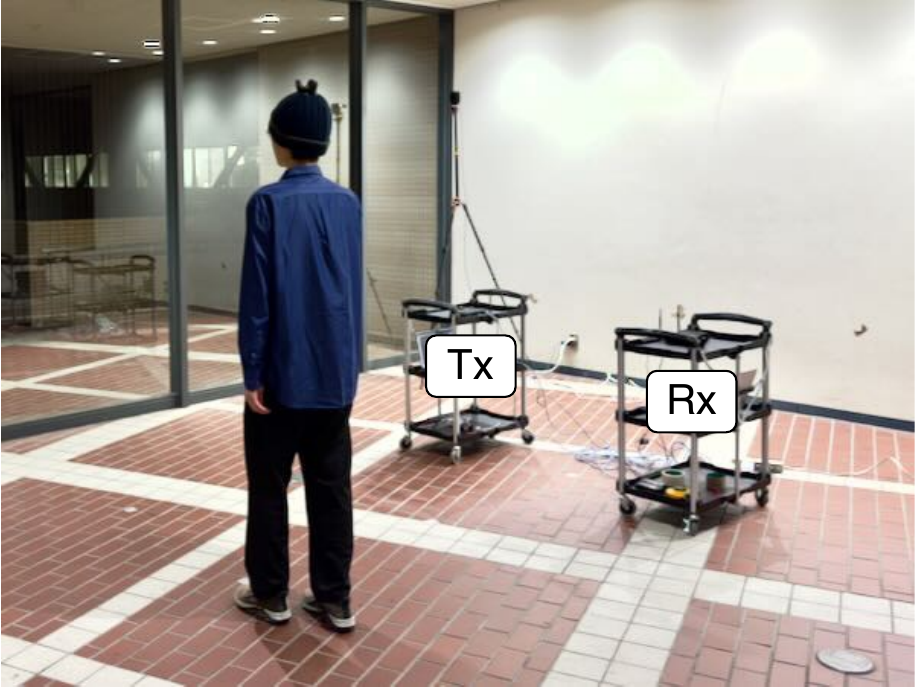}
 \label{fig:moving-human-target}
 }
 \caption{Experimental setup for the human gait experiment.}
 \label{fig:setup}
\end{figure}

The Tx and Rx were placed \SI{1.0}{\meter} apart in a \gls{los} configuration,
corresponding to a reference delay of
\( \tau_{\mathrm{ref}} = \SI{1.0}{\meter}/c = \SI{3.3}{\nano\second} \),
with both antennas positioned at a height of 
\SI{0.8}{\meter} above the floor.
The experimental setup and the straight-line walking trajectory
representing back-and-forth translational gait motion of the human target
were illustrated in Fig.~\ref{fig:setup}.
This simple gait scenario, involving back-and-forth translational motion along
a straight-line trajectory, allows the time-varying CIR and its Doppler characteristics to be examined in an interpretable manner, which will be discussed later in Section~\ref{sec:delay-doppler-estimation}.

\begin{figure}[t]
 \includegraphics[width=\linewidth]{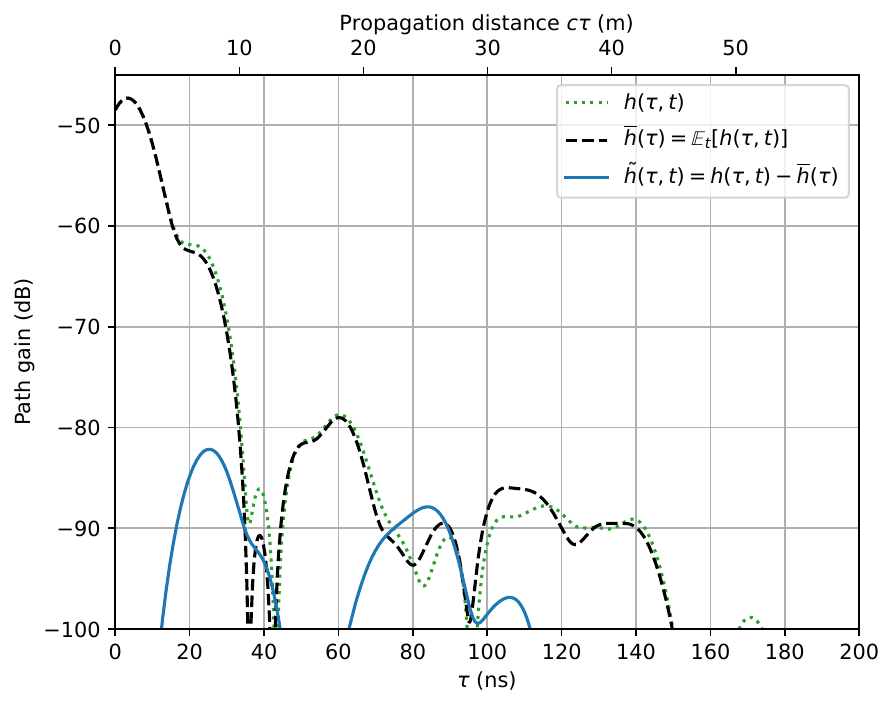}
 \caption{%Snapshot of the observed CIR \( h(\tau,t) \), the estimated static component \( \overline{h}(\tau) \), and the residual CIR \( \tilde{h}(\tau,t) \) at \( t = \SI{12.4}{\second} \).
 Snapshot of the observed CIR \( h(\tau,t) \), the estimated clutter \( \overline{h}(\tau) \), and the resulting residual CIR \( \tilde{h}(\tau,t) \) at \( t = \SI{3.2}{\second} \), where \( h(\tau,t) \) and \( \tilde{h}(\tau,t) \) are averaged over 10 measurements.
 }
 %\caption{10回の平均 □レーダ方程式から何か説明？}
 \label{fig:cir_null}
\end{figure}

\begin{figure}[t]
 \subfloat[][Observed calibrated CIR \( h(\tau,t) \), which is dominated by strong static multipath components.]{%
 %\subfloat[][Observed CIR \( h(\tau,t) \), where strong static multipath components dominate.]{
 %\includegraphics[width=\linewidth]{/Users/kyamamot/Dropbox/due/240804_pico/250727_memory_map/251227_椎野_直線往復/figs/range-time-map_before-static_component_removal.pdf}
 \includegraphics[width=\linewidth]{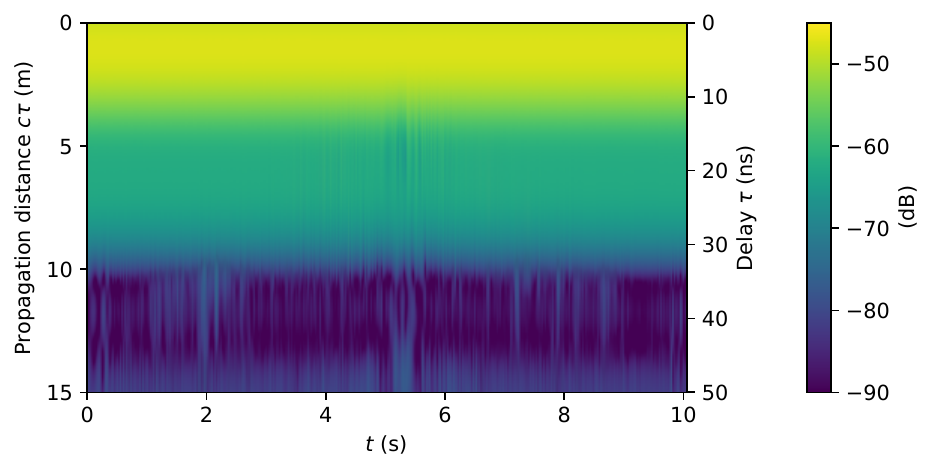}
 \label{fig:h1_tau-t}
 }
 
 \subfloat[][Residual CIR \( \tilde{h}(\tau,t) \) after clutter removal, highlighting the time-varying components induced by human motion.]{
 \includegraphics[width=\linewidth]{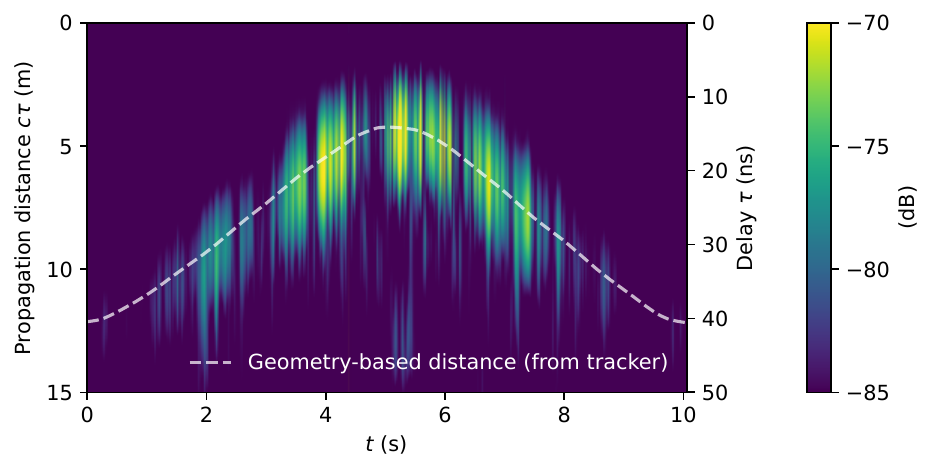}
 \label{fig:h_tau-t}
 }
 \caption{Heatmap representations of the CIR in the delay--time domain measured in the presence of a moving human target. 
 Note that different color scale limits are used in (a) and (b), as the residual CIR in (b) has significantly lower power than the observed CIR in (a).
 }
 \label{fig:h_tau-t_h1_tau-t}
\end{figure}

Fig.~\ref{fig:cir_null} shows a snapshot of the calibrated CIR \( h(\tau,t) \)
at a representative time instant \( t = \SI{3.2}{\second} \),
whereas Fig.~\ref{fig:h_tau-t_h1_tau-t}\subref{fig:h1_tau-t} presents the
corresponding CIR evolution over the entire observation interval
as a delay--time heatmap.
The delay axis \( \tau \) is converted to propagation distance \( c\tau \),
thereby yielding a range--time map.
In this bistatic configuration, it represents the
\emph{bistatic range} \( d_{\mathrm{Tx\text{--}target\text{--}Rx}} \),
i.e., the sum of the Tx--target and target--Rx path lengths
\cite{willis2005bistatic}.

%For visualization purposes, the delay axis \( \tau \) is converted to
%propagation distance \( c\tau \), yielding a range--time map.
%In the bistatic configuration considered here, this propagation distance
%represents the sum of the Tx--target and target--Rx path lengths.
%Following bistatic radar terminology \cite{willis2005bistatic},
%this quantity is hereafter referred to as the \emph{bistatic range}
%\( d_{\mathrm{Tx\text{--}target\text{--}Rx}} \).

Fig.~\ref{fig:h_tau-t_h1_tau-t}\subref{fig:h1_tau-t} shows that the strongest signal
was consistently aligned at \(\tau_{\mathrm{ref}} = \SI{3.3}{\nano\second}\) over time as a result of the LoSRef-enabled delay calibration.
Meanwhile, the time-varying components induced by human motion are not clearly
distinguishable in the raw CIR representation.
Accordingly, clutter removal described in
Section~\ref{ssec:Static-Component-Removal} is applied to obtain the
residual CIR, whose delay--time response is shown in
Fig.~\ref{fig:h_tau-t_h1_tau-t}\subref{fig:h_tau-t}, with the
geometry-based bistatic range
\( d_{\mathrm{Tx\text{--}target\text{--}Rx}} \)
overlaid for reference.
%Accordingly, clutter removal described in
%Section~\ref{ssec:Static-Component-Removal} is applied to obtain the
%residual CIR.
%Fig.~\ref{fig:h_tau-t_h1_tau-t}\subref{fig:h_tau-t} shows the resulting
%delay--time response, with the geometry-based bistatic range
%\( d_{\mathrm{Tx\text{--}target\text{--}Rx}} \)
%overlaid for reference.
A good agreement was observed between the two.
In addition, a snapshot of the residual CIR and the estimated clutter \( \overline{h}(\tau) \) are shown in Fig.~\ref{fig:cir_null}, together with the previously presented results.
Although the human-induced scattering component around \( \tau = \SI{25}{\nano\second} \) is about \SI{20}{\decibel} weaker than the dominant clutter component \( \overline{h}(\tau) \), the residual CIR \( \tilde{h}(\tau,t) \) clearly captured the corresponding temporal variations through clutter removal.

\subsubsection{Short-Time Phase Evolution and Doppler Interpretation}

\begin{figure}[t]
 \centering
 \includegraphics[width=\linewidth]{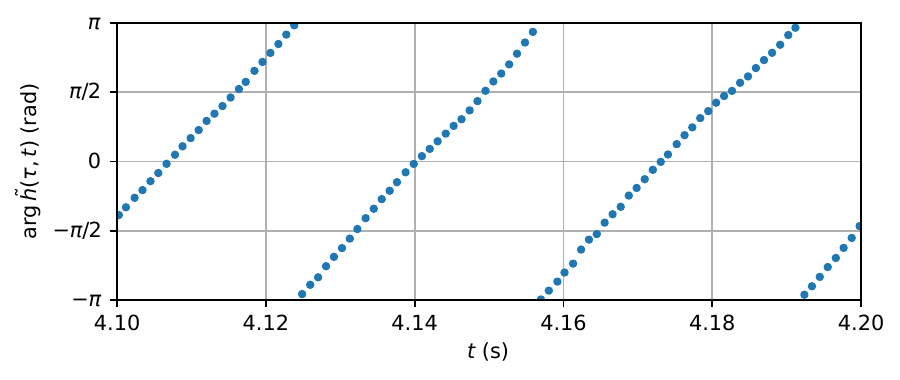}
 \caption{Phase of residual CIR \( \tilde{h}(\SI{20}{\nano\second},t) \) showing continuous phase evolution induced by human walking motion.}
 \label{fig:phase_t_gait}
\end{figure}

To further investigate fine-grained dynamics that are not visible at the
longer time scales above, we focus on the short-time phase
evolution of the residual CIR.
Fig.~\ref{fig:phase_t_gait} shows a zoomed-in view of the phase
$\arg \tilde{h}(\SI{20}{\nano\second},t)$.
%restricted to a duration of \SI{0.1}{\second}.

As observed in Fig.~\ref{fig:phase_t_gait}, the phase advanced by approximately
$2\pi$ every $T=\SI{33}{\milli\second}$, exhibiting a periodic phase evolution.
This indicated that, %over this short interval,
the residual CIR at
$\tau=\SI{20}{\nano\second}$ was dominated by a single complex sinusoidal
component with period $T$.
Accordingly, $\tilde{h}(\SI{20}{\nano\second},t)$ can be locally approximated as
\(
\tilde{h}(\SI{20}{\nano\second},t) \propto \mathrm{e}^{\mathrm{j}2\pi \nu t}
\),
where the frequency
\(
\nu \coloneqq 1/T = \SI{30}{\hertz}
\)
represents the Doppler-induced frequency offset.

\subsubsection{Physical Interpretation in Terms of Bistatic Range Rate}

This periodic phase evolution admits an interpretation in
terms of bistatic range variation.
A phase advance of $2\pi$ corresponds to a reduction of the bistatic range by
one wavelength $\lambda$ over the same interval $T$.
Defining the \emph{bistatic range rate} as the time derivative of the bistatic
range,
\(
\dot{d}_{\mathrm{Tx\text{--}target\text{--}Rx}},
\)
the observed phase evolution implies
\(
\dot{d}_{\mathrm{Tx\text{--}target\text{--}Rx}} = -\lambda / T.
\)
This relationship is consistent with the bistatic Doppler relation
\cite{willis2005bistatic}
\begin{align}
 \nu = - \dot{d}_{\mathrm{Tx\text{--}target\text{--}Rx}} / \lambda.
 \label{eq:def_doppler}
\end{align}

In the example, the positive phase slope indicates a decreasing
bistatic range, corresponding to the target approaching the Tx--Rx pair.
Conversely, when the target recedes, the phase slope---the Doppler
frequency $\nu$---becomes negative.
Therefore, the sign of $\nu$ reveals the direction of motion.

Estimating the Doppler frequency $\nu$ thus corresponds to estimating the
frequency of the underlying complex sinusoidal component
$\ee^{\jj 2\pi \nu t}$ discussed above.
This estimation is obtained via a delay--Doppler representation of the CIR
through Fourier analysis, which generalizes the above short-time analysis,
as described in Section~\ref{sec:delay-doppler-estimation}.

Such fine-grained, phase-coherent delay-domain observations have
traditionally required dedicated channel sounders or radar systems.
In contrast, the present results demonstrate that this regime can be
reached using low-cost COTS \mbox{Wi-Fi} devices.
This shift is enabled by the recent availability of wideband \mbox{Wi-Fi}
signals with bandwidths of up to \SI{160}{\mega\hertz}, together with
preprocessing that allows sufficiently accurate CIR estimation.

\subsubsection{Comparison with Magnitude-Based Wi-Fi Sensing}

\begin{figure}[t]
 \centering
 \includegraphics[width=\linewidth]{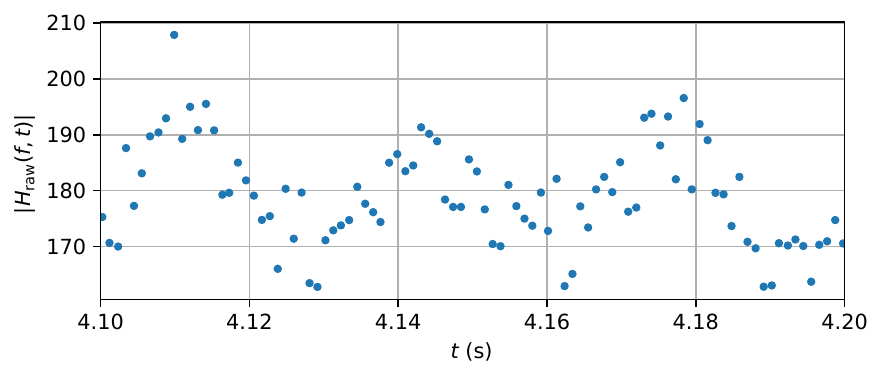}
 \caption{Example of a \mbox{Wi-Fi} sensing signal:
$\lvert H_{\mathrm{raw}}(100\varDelta f,t) \rvert$.}
 \label{fig:Hraw_t_gait}
\end{figure}

To clarify the advantage of this approch,
we now contrast it with conventional magnitude-based sensing.
In these approaches, temporal analysis is performed on the magnitude of the CFR,
$\lvert H_{\mathrm{raw}}(f,t) \rvert$.
%Since the CFR consists of multiple subcarriers,
To reduce dimensionality, 
representative subcarriers are selected (e.g., \cite{Wang2016UbiCompRespiration}) 
\gls{pca} is applied (e.g., \cite{Wang2016UbiCompGait}).
%
%As a representative example,
Fig.~\ref{fig:Hraw_t_gait} shows
$\lvert H_{\mathrm{raw}}(f,t) \rvert$ at a subcarrier
($k=100$) over the same interval as Fig.~\ref{fig:phase_t_gait}.
A periodic fluctuation with a period of approximately
\SI{33}{\milli\second} was also observed.

%To clarify the advantage of this phase-coherent interpretation,
%we now contrast it with conventional magnitude-based \mbox{Wi-Fi} sensing.
%In such approaches, temporal analysis is typically performed on the magnitude
%of the CFR, often after subcarrier selection or
%\gls{pca}-based dimensionality reduction.
%As a representative example, Fig.~\ref{fig:Hraw_t_gait} shows
%$\lvert H_{\mathrm{raw}}(f,t) \rvert$ at a representative subcarrier
%($k=100$) over the same interval as Fig.~\ref{fig:phase_t_gait}.
%A periodic fluctuation with a period of approximately
%\SI{33}{\milli\second} was also observed.

Unlike the phase evolution, however, these magnitude fluctuations
arise from time-varying interference
between human-induced scattering and static multipath components and do not preserve carrier-phase information.
Consequently, while the fluctuation rate reflects motion speed,
the Doppler sign---whether the target is approaching or
receding---cannot be inferred.

\subsection{Observation of Respiration-Induced Body-Surface Motion}
\label{ssec:Respiration}

\subsubsection{Experimental Setup and Range--Time Representation}

To evaluate bistatic range variations that are more gradual than those induced by human gait, we conducted the following respiration experiment.
Although the experiment was performed in a different room from the human gait experiment, the same Tx--Rx geometry and coordinate system as shown in Fig.~\ref{fig:setup}\subref{fig:geometry} were adopted.
The human target was seated on a chair at the position $(\SI{3}{\meter}, \SI{0}{\meter})$, facing the Tx--Rx pair, and performed controlled breathing at a rate of \num{15} breaths per minute, paced by a metronome.
Under this configuration, the bistatic range is
\( d_{\mathrm{Tx\text{--}target\text{--}Rx}} = \SI{6.1}{\meter} \),
corresponding to a bistatic propagation delay of
\( d_{\mathrm{Tx\text{--}target\text{--}Rx}}/c = \SI{20}{\nano\second} \).

\begin{figure}[t]
 \subfloat[][Delay--time representation, where hue denotes the phase and brightness indicates the magnitude in a logarithmic scale.]{%
 \includegraphics[width=\linewidth]{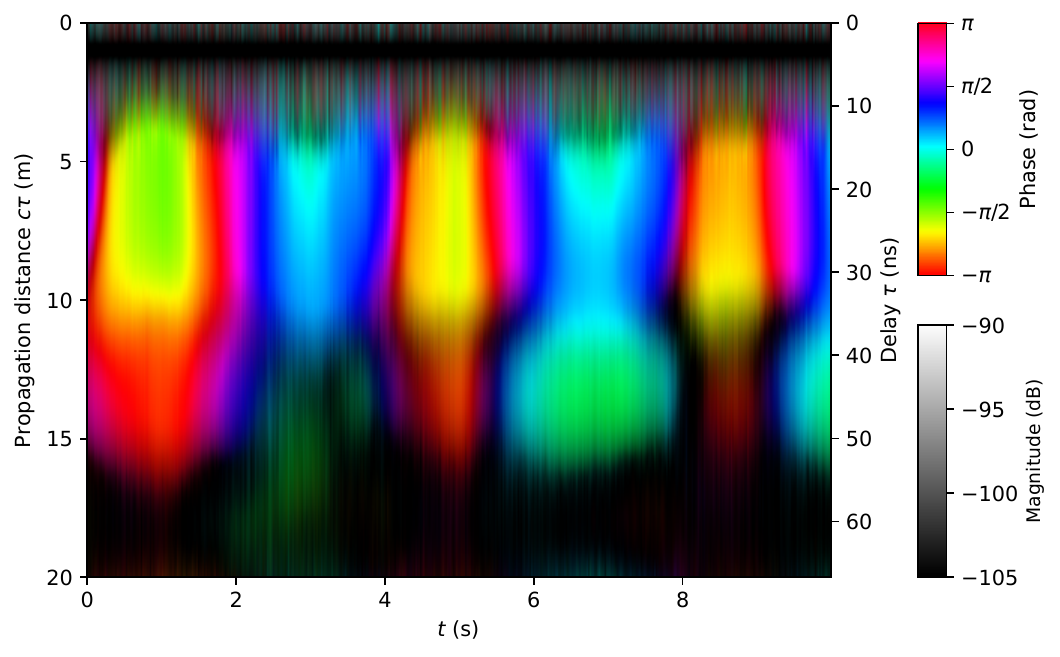}
 \label{fig:phase_tau_t}
 }

 \subfloat[][Phase at $\tau=\SI{20}{\nano\second}$, revealing target-induced variations. For visualization, a constant phase offset is applied to keep the wrapped phase continuous within \( \lbrack -\pi,\pi \rbrack \).]{%
 \includegraphics[width=\linewidth]{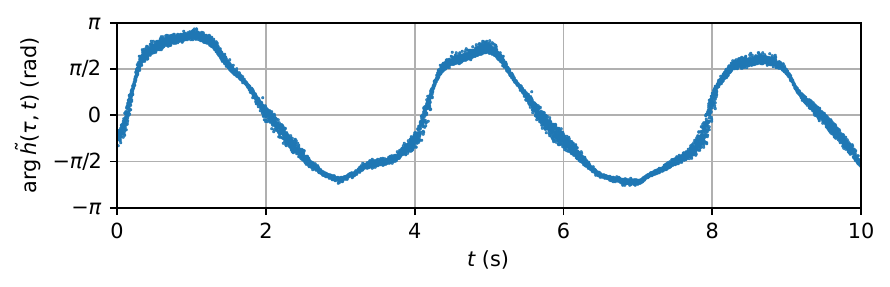}
 \label{fig:phase_t}
 }

 \caption{Residual CIR \( \tilde{h}(\tau,t) \) in a respiration experiment.}
 \label{fig:phase}
\end{figure}

Fig.~\ref{fig:phase}\subref{fig:phase_tau_t} presents a delay--time representation of \( \tilde{h}(\tau,t) \).
In this visualization, the phase of the residual CIR is shown by color, while the magnitude is represented by brightness on a logarithmic scale.
A clear periodic pattern was observed around the geometry-based
bistatic delay of \SI{20}{\nano\second},
suggesting respiration-induced motion of the human target.
Note that this response exhibited a spread in the delay domain.
This spread arises from the finite delay resolution,
\( 1/B = \SI{6.3}{\nano\second} \).

\subsubsection{Phase Evolution and Bistatic Range Interpretation}

The phase at \( \tau = \SI{20}{\nano\second} \) was further shown in Fig.~\ref{fig:phase}\subref{fig:phase_t}.
A clear periodic phase variation with a period of approximately \SI{4}{\second}
was observed, consistent with the controlled respiration rate of
\num{15} breaths per minute.
The peak-to-peak phase excursion was approximately $3\pi/2$.

To interpret this phase variation in terms of physical motion,
let $\varDelta d_\text{Tx--target--Rx}$ denote the respiration-induced variation in the bistatic range $ d_\text{Tx--target--Rx} $.
From the relationship \( (2\pi/\lambda)\,\varDelta d_\text{Tx--target--Rx} = 3\pi/2 \),
the corresponding path-length variation is estimated as
\( \varDelta d_\text{Tx--target--Rx} = 3\lambda/4 = \SI{40}{\milli\meter} \).
This value was reasonable for respiration-induced body-surface motion.
Note that this estimated displacement does not necessarily correspond to a specific anatomical point, but rather reflects respiration-induced variations of dominant reflecting surfaces on the human body.

\subsubsection{Comparison with Magnitude-Based Wi-Fi Sensing}

It is also worth highlighting the difference between the proposed approach and commonly adopted \mbox{Wi-Fi} sensing methods.
In many studies \cite{Wang2016UbiCompRespiration,Kanda2022CCNCCSI,Kanda2025Access}, sensing is performed on the magnitude of CFR, \( \lvert H_{\mathrm{raw}}(f,t) \rvert \).
Since the CFR spans multiple subcarriers, dimensionality reduction---such as subcarrier selection or PCA---is first applied, and the respiration rate is then estimated by performing an FFT along the time axis of the resulting magnitude time series.

\begin{figure}[t]
 \centering
 \subfloat[][CFR magnitude, \( \lvert H_{\mathrm{raw}}(f,t) \rvert \), $k=100$.]{%
 \includegraphics[width=\linewidth]{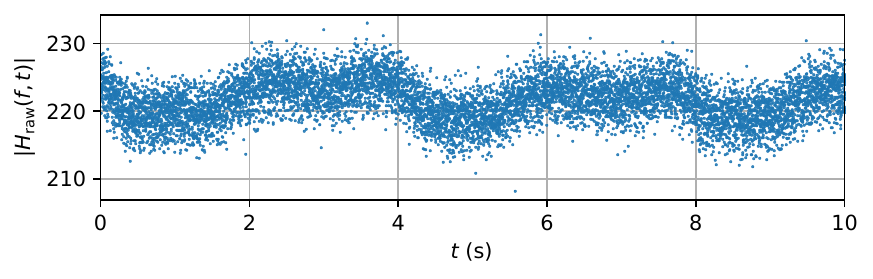}
% \label{fig:Hraw_t_respiration}
 }
 
 \subfloat[][PC2 score of the magnitude of the CFR.]{%
  \includegraphics[width=\linewidth]{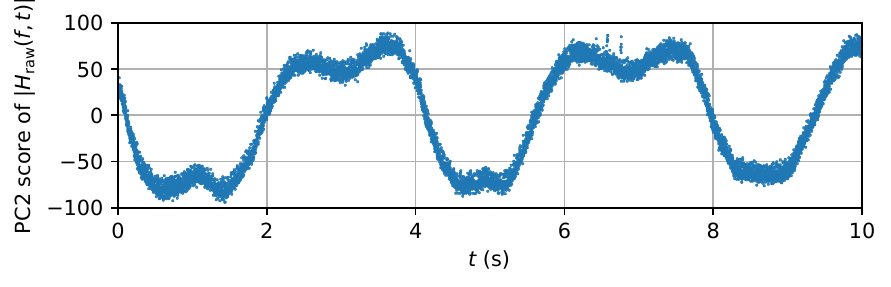}
% \label{fig:PC2score_t}
}
 \caption{Representative examples of magnitude-based \mbox{Wi-Fi} sensing signals.}
 \label{fig:CFR_mag_respiratory}
\end{figure}

%\begin{figure}[t]
% \centering
% %\includegraphics[width=\linewidth]{/Users/kyamamot/Dropbox/due/240804_pico/250727_memory_map/251023_才木/figs/PC3score_t.pdf}
% \caption{PC2 score of the magnitude of the CFR, \( \lvert H_{\mathrm{raw}}(f,t) \rvert \).}
%\end{figure}

As representative examples, the time series of $\lvert H_{\mathrm{raw}}(f,t) \rvert$ at a representative subcarrier ($k=100$) and the PC2 score extracted from
$\lvert H_{\mathrm{raw}}(f,t) \rvert$ are shown in Fig.~\ref{fig:CFR_mag_respiratory}. %~\ref{fig:Hraw_t_respiration} and \ref{fig:PC2score_t}, respectively.
A clear periodic fluctuation was observed, indicating that
respiration-induced motion is captured.
In this sense, applying an FFT to these signals is sufficient for estimating
the respiration rate, and such processing has been adopted in
existing \mbox{Wi-Fi} sensing studies.

However, compared with Fig.~\ref{fig:phase}\subref{fig:phase_t}, these results
exhibited a qualitatively different waveform with substantially larger
noise-like fluctuations in magnitude.
This is because these results are based on the magnitude of the CSI, which inherently mixes respiration-induced motion with other sources of magnitude variation.
Moreover, magnitude-based processing does not preserve the phase
corresponding to the bistatic range itself and therefore lacks a 
physical interpretation.
As a result, the component associated with respiration is often identified
heuristically and varies across devices and environments.
In particular, PC1 frequently reflects dominant static or
hardware-related variations rather than motion-induced dynamics, as
reported in prior studies
\cite{Wang2016UbiCompGait,Wang2016UbiCompRespiration}.

In contrast, the LoSRef-based method exploits the temporal
evolution of the phase, which directly reflects bistatic range variations.
Unlike magnitude-based sensing, this phase-based approach preserves
propagation-induced information and is less sensitive to hardware gain fluctuations.

\section{Delay--Doppler Response Estimation of Moving Human Target}

\label{sec:delay-doppler-estimation}

%\section{Handling Irregular Temporal Sampling for Doppler Estimation in Wi-Fi Systems}
%\label{sec:Time-Domain-Preprocessing}

This section investigates delay--Doppler response estimation of a moving human
target using phase-coherent \mbox{Wi-Fi} CIRs.
We first characterize the impact of irregular temporal sampling in
practical \mbox{Wi-Fi} systems and introduce time-domain resampling.
We then construct a delay--Doppler representation and demonstrate
range--Doppler estimation through experiments.

%In practical \mbox{Wi-Fi} systems, CFR measurements are not acquired on a perfectly
%uniform time grid.
%This section experimentally characterized the resulting inter-packet timing
%variability and introduced time-domain resampling to enable
%reliable phase-coherent Doppler estimation.

\subsection{Inter-Packet Timing Variations in Contention-Based Wi-Fi}

Although packet transmissions are scheduled at regular intervals,
the actual observation times %at the Rx
are not guaranteed
to be uniformly spaced.
This is due to contention-based medium access, missed receptions, and the removal of outliers.

\begin{figure}[t]
 %\subfloat[][Temporal variation of the observed inter-packet intervals over the measurement duration.]{%
 %\includegraphics[width=.45\linewidth]{/Users/kyamamot/Dropbox/due/240804_pico/250727_memory_map/251227_椎野_直線往復/figs/intersample_time.pdf}%
 %\includegraphics[width=.45\linewidth]{intersample_time.pdf}%
 %\label{fig:intersample_time}
 %}
 %\hfill
 %\subfloat[][Empirical CDF of the observed inter-packet intervals.]{%
 %\includegraphics[width=.47\linewidth]{/Users/kyamamot/Dropbox/due/240804_pico/250727_memory_map/251227_椎野_直線往復/figs/intersample_time_cdf.pdf}%
 \includegraphics[width=\linewidth]{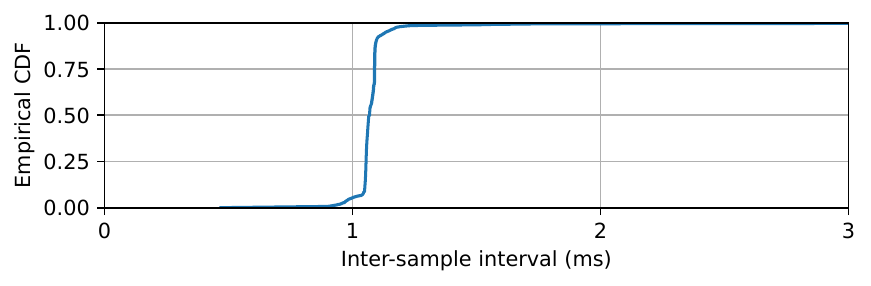}%
 %}
 \caption{Empirical CDF of the observed inter-packet intervals.
 %Temporal variation and empirical distribution of the observed inter-packet intervals.
 Although the nominal sampling interval is \SI{1}{\milli\second}, the intervals exhibit noticeable fluctuations, with a median of \SI{1.069}{\milli\second}.}
 \label{fig:intersample_time_cdf}
 %\label{fig:intersample}
\end{figure}

%This timing variability was examined in Fig.~\ref{fig:intersample}.
%Fig.~\ref{fig:intersample}\subref{fig:intersample_time} illustrates the inter-packet intervals computed from the measurements in the experiment described in Section~\ref{ssec:range-estimation}, where packet transmissions were intended to follow a \SI{1}{\milli\second} interval.
%The inter-packet intervals were spread over a range from approximately \SI{0.5}{\milli\second} to \SI{6.5}{\milli\second}, indicating deviations from the nominal sampling period.

Fig.~\ref{fig:intersample_time_cdf} shows the empirical cumulative distribution function (CDF) of the inter-packet intervals
obtained in the experiment described in Section~\ref{ssec:range-estimation}, where packet transmissions were intended to follow a \SI{1}{\milli\second} interval.
The observed intervals exhibited a median of \SI{1.069}{\milli\second}, with a median absolute deviation of \SI{0.017}{\milli\second}, indicating small but non-negligible timing jitter around the nominal sampling interval.

\subsection{Uniform Temporal Resampling for Phase-Coherent Doppler Estimation}
\label{ssec:Uniform-Resampling}

Since Doppler estimation generally requires uniformly spaced observations in time,
a uniform time grid is constructed, where the grid spacing
$\varDelta t$ is set to the median inter-packet interval,
i.e., $\varDelta t = \SI{1.069}{\milli\second}$ in the present measurements.
Each observed CIR snapshot is first associated with its temporally nearest grid point.
The remaining grid points are then filled by interpolation:
the magnitude is interpolated linearly, while the phase is interpolated
after phase unwrapping so as to preserve temporal continuity.
Although more advanced treatments of non-uniform sampling are possible, the above
procedure is sufficient for accurate phase-coherent Doppler estimation in this work.

%allowing the same observation to be assigned to multiple grid points.
%\footnote{More sophisticated approaches include interpolating the measurements to arbitrary time instants or applying non-uniform discrete Fourier transform (NUDFT). %
%These alternatives were not pursued in this study, as the proposed method already yields reliable Doppler frequency estimates. % 
%Note that NUDFT has also been employed in Chronos \cite{Vasisht2016NSDI}; however, it is applied to the frequency-to-delay transform, rather than to the time-to-Doppler domain considered here.}
% Chronosはfrequency -> delayで使っていて，我々はtime -> Dopplerで使っている

\subsection{Delay--Doppler Representation and Doppler Interpretation}

The delay--Doppler representation is obtained by Fourier transforming the
CIR $h(\tau,t)$ with respect to time $t$ \cite{Bello1963,Molisch2023}.
While Section~\ref{ssec:range-estimation} illustrated Doppler behavior over a short time
interval at a fixed delay, the delay--Doppler representation generalizes this
observation by characterizing Doppler frequency as a function of delay.
In practice, we apply an \gls{stft} along the time axis to the
phase-coherent and uniformly resampled CIR \( h(\tau,t) \), yielding a time-varying
delay--Doppler representation $s(\tau,\nu;t)$, where $t$ denotes the
center time of the analysis window.\footnote{The CIR $h(\tau,t)$ may be taken either before or after clutter removal, since clutter components are suppressed on a per-STFT-segment basis.}
This representation jointly captures the temporal evolution of multipath
delays and their associated Doppler frequencies.

Although the experiment employs a bistatic configuration, the Tx--Rx
separation is small enough that the geometry can be approximated as
monostatic. In this limit, $\nu=-2v/\lambda$, where $v$ denotes the
target radial velocity. Accordingly, we interpret the Doppler frequency
using the \textit{effective radial velocity} $-\nu\lambda/2$.

%Although the experiment employs a bistatic configuration, the Tx--Rx
%separation is small enough that the geometry can be approximated as
%monostatic. In this limit, the bistatic range rate satisfies
%$\dot{d}_{\text{Tx--target--Rx}} = 2v$, where $v$ denotes the
%radial velocity of the target. Substituting this relation into
%(\ref{eq:def_doppler}) yields $\nu = -2v/\lambda$.
%For convenience, we therefore interpret the Doppler frequency using the
%\textit{effective radial velocity} $-\nu\lambda/2$ in the following discussion.

%Although the experiment employs a bistatic configuration, the Tx--Rx
%separation is sufficiently small that the geometry can be approximated
%as monostatic.
%In this limit, the bistatic range
%$d_{\text{Tx--target--Rx}}$ reduces to twice the one-way Tx/Rx--target
%distance, and the bistatic range rate becomes
%\begin{align}
% \dot{d}_{\text{Tx--target--Rx}} = 2v,
% \label{eq:rel_bistatic_range_rate__radial_velocity}
%\end{align}
%where $v$ denotes the radial velocity defined as the time derivative
%of the one-way Tx/Rx--target distance.
%Substituting (\ref{eq:rel_bistatic_range_rate__radial_velocity}) into 
%(\ref{eq:def_doppler}) yields
%\begin{align}
% \nu = -{2v}/{\lambda}.
% \label{eq:doppler-velocity-mono}
%\end{align}
%For ease of interpretation, we therefore employ an \emph{effective radial
%velocity} scaled as \( -\nu\lambda/2 \) in the following discussion,
%instead of the bistatic range rate.

\subsection{Experimental Results: Range--Doppler Estimation}
\label{ssec:Range--Doppler-Estimation}
%\subsection{Experimental Results: Range--Doppler and Doppler--Time Analysis}

\begin{table}[t]
 \centering
 \caption{STFT Parameters for Doppler Analysis}
 \begin{tabular}{cc}
  \toprule
  STFT segment length & \num{256} \\
  STFT overlap        & \num{224} \\
  Doppler-domain interpolation rate & \num{8} \\
  STFT window & Hann \\
  \bottomrule
 \end{tabular}
 \label{tab:STFTParameters}
\end{table}

The main STFT parameters are summarized in Table~\ref{tab:STFTParameters}.
Similar to the Fourier transform from the frequency domain to the delay domain in \eqref{eq:H2h}, a Hann window was applied to each STFT segment of length \num{256} samples to suppress spectral leakage,
with an overlap of \num{224} samples (\num{87.5}\%), corresponding to a hop size of \SI{32}{\milli\second}.
With an STFT segment of \num{256} samples and a time grid of
$\varDelta t \approx \SI{1}{\milli\second}$, the resulting Doppler frequency
resolution, determined by the effective STFT window duration, is approximately $\varDelta \nu \approx \SI{3.9}{\hertz}$,
which is sufficient to resolve human motion-induced Doppler components.
In addition, Doppler-domain oversampling with a factor of \num{8} was performed by zero-padding in the time domain.
Furthermore, the mean value of each STFT segment was removed to mitigate DC components.
This operation %can be interpreted
corresponds to applying clutter removal described in Section~\ref{ssec:Static-Component-Removal} at the level of individual STFT segments.

%The Doppler spectrum is evaluated over both positive and negative frequencies to distinguish approaching and receding motions.

\begin{figure}[t]
 \includegraphics[width=\linewidth]{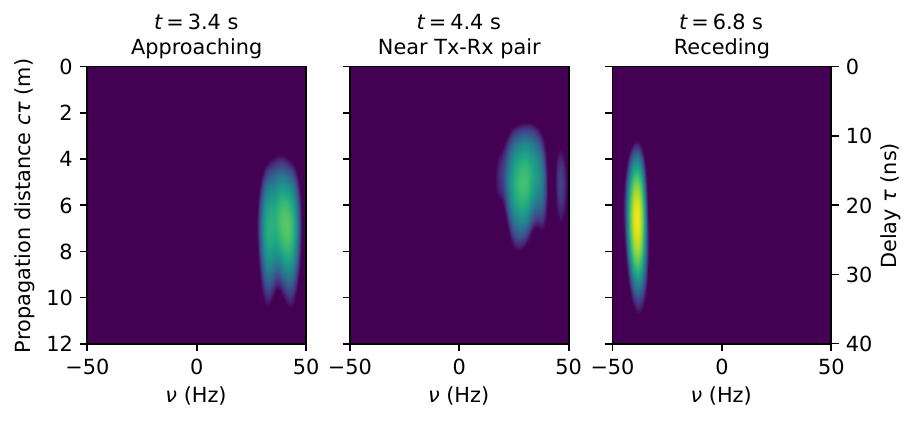}
 \caption{Representative snapshots of \( {s}(\tau,\nu; t) \).
As the target passes near the Tx--Rx pair, the Doppler frequency changes sign
from positive to negative, and the propagation delay reaches a minimum
around this passage.}
% \caption{Representative snapshots of the range--Doppler representation
%\( {s}(\tau,\nu; t) \) obtained from the moving human target experiment.
%As the target passes near the Tx--Rx pair, the Doppler frequency changes sign from positive to negative, and the delay reaches its minimum at the closest approach.
% }
 \label{fig:range_doppler}
\end{figure}

\begin{figure}[t]
 \centering
 
 \subfloat[][Estimated bistatic range $c\tau^\star$.]{%
 \includegraphics[width=.9\linewidth]{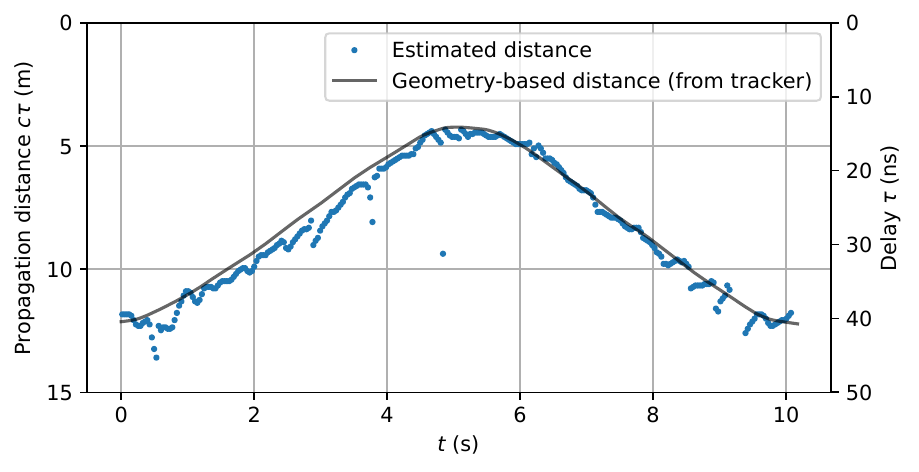}%
 \label{fig:range-time-peak}
 }
 
 \subfloat[][Estimated Doppler frequency $\nu^\star$.]{%
 \includegraphics[width=.9\linewidth]{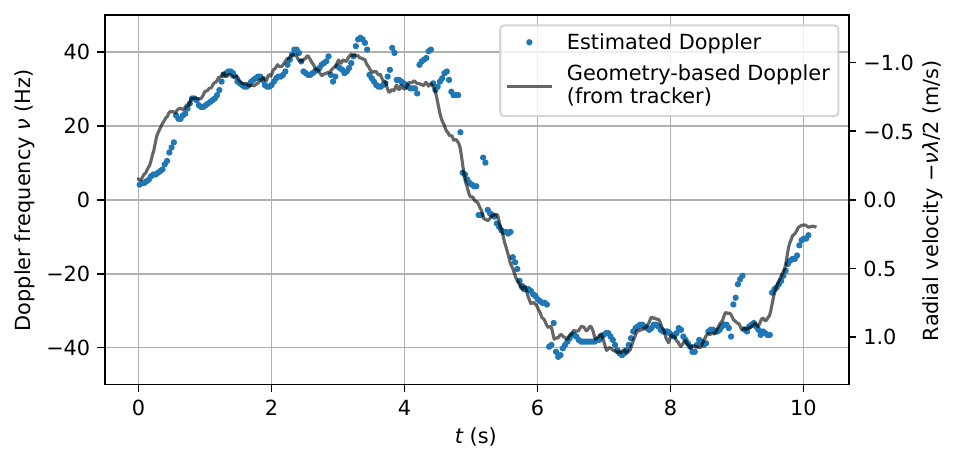}%
 \label{fig:doppler-time-peak}
 }
 \caption{Estimated bistatic range $c\tau^\star$ and Doppler frequency $\nu^\star$ of a moving human target, extracted by selecting the maximum of $s(\tau,\nu;t)$ at each time $t$.
 The curve indicates the geometry-based value derived from the motion-tracking system, shown for reference.}
 \label{fig:peak}
\end{figure}

A representative example of the range--Doppler estimation \( {s}(\tau,\nu; t) \), obtained from the moving human target measurements described in Section~\ref{ssec:range-estimation}, was shown in Fig.~\ref{fig:range_doppler}.
The evolution of the delay--Doppler structure over time reflected the target motion relative to the Tx--Rx pair.
Fig.~\ref{fig:peak} shows the estimated bistatic range $c\tau^\star$ and
Doppler frequency $\nu^\star$ of the moving human target,
together with the corresponding geometry-based values derived from the
motion-tracking system.
Here, $\tau^\star$ and $\nu^\star$ were obtained by selecting, at each time instant $t$,
the maximum of the delay--Doppler representation $s(\tau,\nu;t)$, i.e.,
\begin{align}
 (\tau^\star, \nu^\star) \coloneqq 
 \arg\max_{(\tau,\nu)} s(\tau,\nu;t).
 \label{eq:tau_nu_star}
\end{align}
Overall, the estimated bistatic range and Doppler frequency %exhibited trends that
are broadly consistent with the geometry-based reference.
It should be noted that the tracking system captures the motion of a marker
attached to the target's head, while the Doppler estimates may be influenced
by distributed scattering from multiple body parts.
Therefore, perfect agreement is not expected.
More specifically, Fig.~\ref{fig:peak}\subref{fig:range-time-peak}
corresponds to the delay--time heatmap in
Fig.~\ref{fig:h_tau-t_h1_tau-t}\subref{fig:h_tau-t}.
The bistatic range shown here is given by $c\tau^\star$ extracted from the
delay--Doppler representation, rather than the peak of
the residual CIR $\tilde{h}(\tau,t)$.
%The bistatic range shown here is obtained as $c\tau^\star$
%rather than being obtained by directly taking the peak of the residual CIR
%$\tilde{h}(\tau,t)$.
%where $\tau^\star$ is defined in \eqref{eq:tau_nu_star},

In Fig.~\ref{fig:peak}\subref{fig:doppler-time-peak}, the estimated Doppler frequency $\nu^\star$
is shown together with the effective radial velocity
$-\nu^\star\lambda/2$.
For typical \mbox{Wi-Fi} carrier frequencies, human motion gives rise to Doppler
frequencies on the order of several tens of hertz
\cite{Wang2016UbiCompGait,Xu2017IMWUT}.
With a uniform time grid of $\varDelta t \approx \SI{1}{\milli\second}$,
the resulting effective Doppler bandwidth of $\pm\SI{500}{\hertz}$ is
therefore sufficient to capture such motion-induced Doppler components.

%, evaluated using \eqref{eq:doppler-velocity-mono}, on the right axis.
%The absolute Doppler frequency scale is determined not only by the target
%motion but also by the carrier frequency.

\begin{figure}[t]
 \centering
 \includegraphics[width=.9\linewidth]{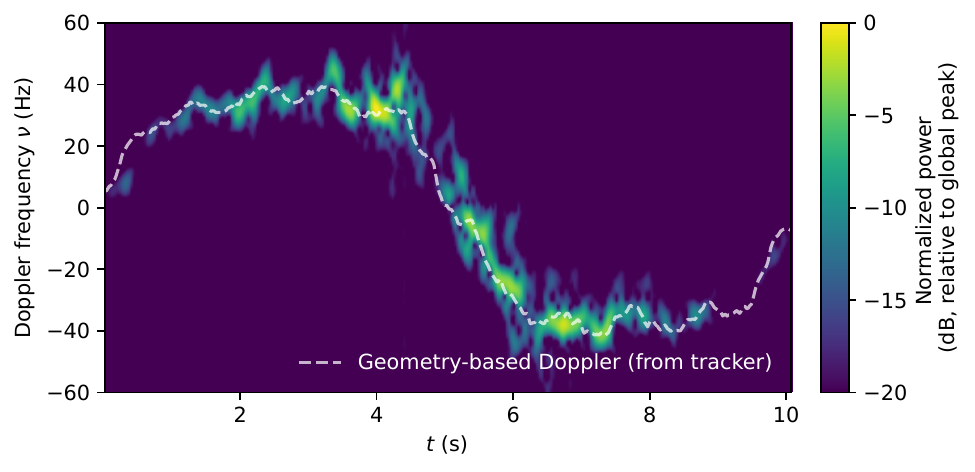}
 \caption{Doppler--time representation obtained by incoherent integration of the delay--Doppler response \( {s}(\tau,\nu; t) \).}
 \label{fig:s_nu-t}
\end{figure}

The peak of $s(\tau,\nu;t)$, i.e., $(\tau^\star, \nu^\star)$, provides
point estimates of the bistatic range and Doppler frequency, but does not
capture the full Doppler energy distribution.
To complement this point-wise representation, we additionally examined a
Doppler--time representation obtained by incoherently integrating the
power $\lvert s(\tau,\nu;t) \rvert^2 $ along the delay axis, as shown in
Fig.~\ref{fig:s_nu-t}.
This representation visualizes the temporal Doppler energy and exhibits a similar trend to that in
Fig.~\ref{fig:peak}\subref{fig:doppler-time-peak};
however, the two representations are not directly comparable, as they are
derived through different aggregation procedures.

\subsection{Comparison with Frequency-Domain Magnitude-Based \mbox{Wi-Fi} Sensing}

For comparison, we present a conventional \mbox{Wi-Fi} sensing result
that operates in the frequency domain and does not rely on the CIR representation.
As discussed in Section~\ref{ssec:Doppler-Based-Wi-Fi-Sensing},
many existing approaches perform sensing directly on the magnitude of the CFR,
$\lvert H_{\mathrm{raw}}(f,t) \rvert$.
%Since the CFR consists of multiple subcarriers, dimensionality reduction is typically required
%prior to temporal analysis.
In this paper, PCA is used as a representative example of
dimensionality reduction, as in \cite{Wang2016UbiCompGait}.
%although alternative strategies such as subcarrier selection
%are also widely adopted
%(e.g., \cite{Wang2016UbiCompRespiration}).

\begin{figure}[t]
 \centering
 \includegraphics[width=.9\linewidth]{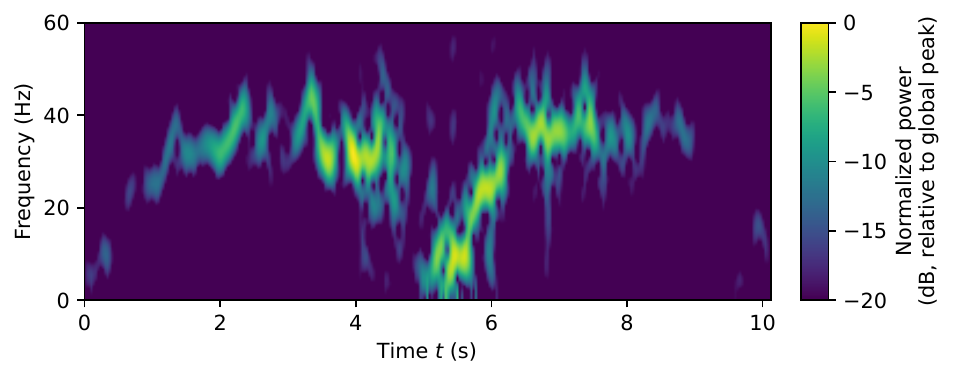}
 \caption{Doppler-related time--frequency representation obtained from the STFT of the PC3 score of \( \lvert H_{\mathrm{raw}}(f,t) \rvert \).
 }
 \label{fig:PCA}
\end{figure}

Among the resulting PC scores, the STFT is applied to the PC3 score,
which exhibits target-induced temporal variations.
It should be noted, however, that the PC capturing motion-induced
dynamics is device- and environment-dependent \cite{Wang2016UbiCompGait,Wang2016UbiCompRespiration},
and its identification relies on empirical or heuristic criteria rather than a physical model.
The resulting Doppler-related time--frequency representation
is shown in Fig.~\ref{fig:PCA}.

A key limitation of this magnitude-based frequency-domain approach is that
it does not yield a physically meaningful Doppler frequency,
because the propagation-induced phase evolution is not preserved,
as explained in Section~\ref{ssec:Doppler-Based-Wi-Fi-Sensing} with reference to \cite{Li2021WCOM}.
As a result, the spectrum in Fig.~\ref{fig:PCA} is symmetric with respect to frequency sign
and collapses positive and negative Doppler components,
making it impossible to distinguish approaching and receding motions.

By contrast, the proposed LoSRef-based approach operates on the complex-valued
residual CIR, whose Doppler spectrum is asymmetric with respect to
frequency sign.
This asymmetry preserves Doppler sign information and enables 
discrimination between approaching and receding motions, which is 
impossible with magnitude-based processing.

%\section{Discussion}
%\label{sec:discussion}
%
%
%The experimental validation was performed using the Intel AX210 chipset.
%While multiple tools for CSI extraction from IEEE 802.11ax devices exist
%\cite{Zhang2025SENSOR,feitcsi},
%they are all restricted to the Intel AX2xx family,
%precluding validation on alternative chipsets.
%Nevertheless, the LoSRef-based framework is chipset-agnostic,
%as the subsequent processing after phase-coherent CIR construction
%follows standard bistatic radar signal processing principles,
%and is therefore applicable to any system that provides a
%complex-valued CFR and a CIR with sufficient delay resolution.

%Nevertheless, the LoSRef-based framework
%is chipset-agnostic and applicable to any system that provides a complex-valued CFR and a CIR with sufficient delay resolution.

%Recent studies have shown that range--Doppler information can be extracted
%from CSI using COTS IEEE 802.11ax \mbox{Wi-Fi} devices under monostatic configurations
%\cite{Sanson2025GLOBECOM}.
%However, extending such capabilities to bistatic scenarios remains challenging,
%primarily due to the lack of a reliable reference signal between the Tx
%and Rx. %, and the reliance on non-public firmware in existing demonstrations.

\section{Conclusion}
\label{sec:Conclusion}

This paper presented a LoSRef-based bistatic
\mbox{Wi-Fi} radar framework that enables phase-coherent delay--Doppler analysis
using only unmodified COTS \mbox{Wi-Fi} devices.
LoSRef leverages the Tx--Rx LoS path as a delay and phase reference;
its temporal stability enables phase-coherent alignment
across successive packet receptions.
As a result, the proposed framework eliminates the need for wired references
or dedicated reference antennas and achieves phase-coherent bistatic radar operation
with a single-antenna, single-port Rx,
thereby breaking a key deployment barrier in existing \mbox{Wi-Fi} radar systems.

Through human gait and respiration experiments, we confirmed that phase-coherent bistatic radar functionality can be 
realized within typically deployed, communication-oriented \mbox{Wi-Fi} systems.
Specifically, the proposed framework enables physically meaningful delay and Doppler estimates,
including gait direction discrimination and respiration-induced sub-wavelength displacement measurements.

The LoSRef-based framework further suggests that
\mbox{Wi-Fi} sensing can move beyond application-specific processing.
Once phase-coherent representations are restored,
radar signal processing techniques can be applied to \mbox{Wi-Fi} sensing,
pointing to research direction for future ISAC system design.

\section*{Acknowledgment}

The author would like to thank \mbox{Mr.~Genichiro~Shiino} 
for providing the experimental data used in this study and for valuable discussions.
The author also thanks \mbox{Prof.~Katsuyuki~Haneda} for his insightful comments
on an early draft of this paper, and \mbox{Prof.~Hiromitsu~Awano} for his valuable comments
on vital sensing experiments.
The author used generative AI tools for language editing during the
preparation of this manuscript. The author reviewed and takes full
responsibility for the final content.

\begin{IEEEbiography}
	[{\includegraphics[width=1in,height=1.25in,clip,keepaspectratio]{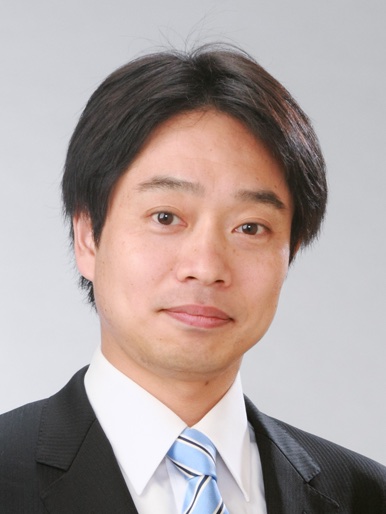}}]
	{Koji Yamamoto}
	(S'03--M'06--SM'20)
   received the B.E.\ degree in electrical and electronic engineering from Kyoto University in 2002, and the master and Ph.D.\ degrees in Informatics from Kyoto University in 2004 and 2005, respectively.
	From 2004 to 2005, he was a research fellow of the Japan Society for the Promotion of Science (JSPS).
	From 2005 to 2023, he was an assistant and associate professor at the Graduate School of Informatics, Kyoto University.
	From 2008 to 2009, he was a visiting researcher at Wireless@KTH, Royal Institute of Technology (KTH) in Sweden.
	Since 2023, he has been a professor with the Faculty of Information and Human Sciences, Kyoto Institute of Technology.
	He serves as an editor of IEEE Open Journal of Vehicular Technology.
 %and Journal of Communications and Information Networks.
	He was a tutorial lecturer in ICC 2019 and a symposium co-chair of GLOBECOM 2021.
	His research interests include radio resource management, game theory, and machine learning.
	He received the PIMRC 2004 Best Student Paper Award in 2004, the Ericsson Young Scientist Award in 2006, IEEE Transactions on Mobile Computing Best Paper Award in 2024.
	He also received the Young Researcher's Award, the Paper Award, SUEMATSU-Yasuharu Award, Educational Service Award, the Paper Award from the IEICE of Japan in 2008, 2011, 2016, 2020, and 2022, respectively, and IEEE Kansai Section GOLD Award in 2012.
	He is a Senior Member of the IEICE and a Member of the Operations Research Society of Japan.
\end{IEEEbiography}

\vfill


\begin{thebibliography}{10}
\providecommand{\url}[1]{#1}
\csname url@samestyle\endcsname
\providecommand{\newblock}{\relax}
\providecommand{\bibinfo}[2]{#2}
\providecommand{\BIBentrySTDinterwordspacing}{\spaceskip=0pt\relax}
\providecommand{\BIBentryALTinterwordstretchfactor}{4}
\providecommand{\BIBentryALTinterwordspacing}{\spaceskip=\fontdimen2\font plus
\BIBentryALTinterwordstretchfactor\fontdimen3\font minus
  \fontdimen4\font\relax}
\providecommand{\BIBforeignlanguage}[2]{{%
\expandafter\ifx\csname l@#1\endcsname\relax
\typeout{** WARNING: IEEEtran.bst: No hyphenation pattern has been}%
\typeout{** loaded for the language `#1'. Using the pattern for}%
\typeout{** the default language instead.}%
\else
\language=\csname l@#1\endcsname
\fi
#2}}
\providecommand{\BIBdecl}{\relax}
\BIBdecl

\bibitem{Liu2022JSAC}
F.~Liu, Y.~Cui, C.~Masouros, J.~Xu, T.~X. Han, Y.~C. Eldar, and S.~Buzzi,
  ``Integrated sensing and communications: Toward dual-functional wireless
  networks for {6G} and beyond,'' \emph{{IEEE} J. Sel. Areas Commun.}, vol.~40,
  no.~6, pp. 1728--1767, Jun. 2022.

\bibitem{Ma2019wifi}
Y.~Ma, G.~Zhou, and S.~Wang, ``{WiFi} sensing with channel state information: A
  survey,'' \emph{ACM Comput. Surv.}, vol.~52, no.~3, pp. 1--36, Jun. 2019.

\bibitem{Tan2022IOT}
S.~Tan, Y.~Ren, J.~Yang, and Y.~Chen, ``Commodity {WiFi} sensing in ten years:
  Status, challenges, and opportunities,'' \emph{{IEEE} Internet Things J.},
  vol.~9, no.~18, pp. 17\,832--17\,843, Apr. 2022.

\bibitem{Chen2023ACM}
C.~Chen, G.~Zhou, and Y.~Lin, ``Cross-domain {WiFi} sensing with channel state
  information: A survey,'' \emph{ACM Comput. Surv.}, vol.~55, no.~11, pp.
  1--37, Feb. 2023.

\bibitem{Moussa2009PERCOM}
M.~Moussa and M.~Youssef, ``Smart devices for smart environments: Device-free
  passive detection in real environments,'' in \emph{Proc. IEEE Int. Conf.
  Pervasive Comput. Commun. (PERCOM)}, Galveston, TX, USA, Mar. 2009, pp. 1--6.

\bibitem{Seifeldin2013TMC}
M.~Seifeldin, A.~Saeed, A.~E. Kosba, A.~El-Keyi, and M.~Youssef, ``Nuzzer: A
  large-scale device-free passive localization system for wireless
  environments,'' \emph{{IEEE} Trans. Mobile Comput.}, vol.~12, no.~7, pp.
  1321--1334, Jul. 2013.

\bibitem{Halperin2011SIGCOMM}
D.~Halperin, W.~Hu, A.~Sheth, and D.~Wetherall, ``Tool release: Gathering
  802.11n traces with channel state information,'' \emph{ACM SIGCOMM Comput.
  Commun. Review (CCR)}, vol.~41, no.~1, p.~53, Jan. 2011.

\bibitem{Du2025CSTO}
R.~Du, H.~Hua, H.~Xie, X.~Song, Z.~Lyu, M.~Hu, Narengerile, Y.~Xin, S.~McCann,
  M.~Montemurro, T.~X. Han, and J.~Xu, ``An overview on {IEEE} 802.11bf: {WLAN}
  sensing,'' \emph{{IEEE} Commun. Surveys Tuts.}, vol.~27, no.~1, pp. 184--217,
  Feb. 2025.

\bibitem{Guo2008Radar}
H.~Guo, K.~Woodbridge, and C.~J. Baker, ``Evaluation of {WiFi} beacon
  transmissions for wireless based passive radar,'' in \emph{Proc. IEEE Radar
  Conf. 2008}, Rome, Italy, May 2008, pp. 1--6.

\bibitem{Colone2012AES}
F.~Colone, P.~Falcone, C.~Bongioanni, and P.~Lombardo, ``{WiFi}-based passive
  bistatic radar: Data processing schemes and experimental results,''
  \emph{{IEEE} Trans. Aerosp. Electron. Syst.}, vol.~48, no.~2, pp. 1061--1079,
  Apr. 2012.

\bibitem{Chetty2012GRS}
K.~Chetty, G.~E. Smith, and K.~Woodbridge, ``Through-the-wall sensing of
  personnel using passive bistatic {WiFi} radar at standoff distances,''
  \emph{{IEEE} Trans. Geosci. Remote Sens.}, vol.~50, no.~4, pp. 1218--1226,
  Apr. 2012.

\bibitem{Falcone2012AES}
P.~Falcone, F.~Colone, and P.~Lombardo, ``Potentialities and challenges of
  {WiFi}-based passive radar,'' \emph{{IEEE} Aerosp. Electron. Syst. Mag.},
  vol.~27, no.~11, pp. 15--26, Nov. 2012.

\bibitem{Pastina2015VT}
D.~Pastina, F.~Colone, T.~Martelli, and P.~Falcone, ``Parasitic exploitation of
  {Wi-Fi} signals for indoor radar surveillance,'' \emph{{IEEE} Trans. Veh.
  Technol.}, vol.~64, no.~4, pp. 1401--1415, Apr. 2015.

\bibitem{Keerativoranan2018}
N.~Keerativoranan, A.~Haniz, K.~Saito, and J.~Takada, ``Mitigation of {CSI}
  temporal phase rotation with {B2B} calibration method for fine-grained motion
  detection analysis on commodity {Wi-Fi} devices,'' \emph{Sensors}, vol.~18,
  no.~11, pp. 1--18, Nov. 2018.

\bibitem{Song2020IOT}
H.~Song, B.~Wei, Q.~Yu, X.~Xiao, and T.~Kikkawa, ``{WiEps}: Measurement of
  dielectric property with commodity {WiFi} device---{An} application to
  ethanol/water mixture,'' \emph{{IEEE} Internet Things J.}, vol.~7, no.~12,
  pp. 11\,667--11\,677, Dec. 2020.

\bibitem{Qiu2023IOT}
J.~Qiu, P.~Zheng, K.~Chi, R.~Xu, and J.~Liu, ``Respiration monitoring in
  high-dynamic environments via combining multiple {WiFi} channels based on
  wire direct connection between {RX/TX},'' \emph{{IEEE} Internet Things J.},
  vol.~10, no.~2, pp. 1558--1573, Jan. 2023.

\bibitem{Yamamoto2025EuCAP}
K.~Yamamoto and K.~Haneda, ``Delay-synchronous wideband channel sounding using
  off-the-shelf multi-antenna {WiFi} devices,'' in \emph{Proc. 19th Eur. Conf.
  Ant. Propag. (EuCAP 2025)}, Stockholm, Sweden, Apr. 2025, pp. 1--5.

\bibitem{ieee80211ax}
{IEEE Std 802.11ax-2021}, ``Wireless {LAN} medium access control ({MAC}) and
  physical layer ({PHY}) specifications, {Amendment}: Enhancements for high
  efficiency {WLAN},'' May 2021.

\bibitem{Wu2016UBICOMP}
D.~Wu, D.~Zhang, C.~Xu, Y.~Wang, and H.~Wang, ``{WiDir}: Walking direction
  estimation using wireless signals,'' in \emph{Proc. ACM UbiComp 2016},
  Heidelberg, Germany, Sep. 2016, pp. 351--362.

\bibitem{Tamai2025TVT}
K.~Tamai, K.~Yamamoto, N.~Kato, and T.~Negishi, ``Device-free pedestrian
  tracking using {CSI} sampled at sub-{Nyquist} rate for human gait,''
  \emph{{IEEE} Trans. Veh. Technol.}, vol.~74, no.~4, pp. 6088--6098, Apr.
  2025.

\bibitem{Kotaru2015SIGCOMM}
M.~Kotaru, K.~Joshi, D.~Bharadia, and S.~Katti, ``{SpotFi}: Decimeter level
  localization using {WiFi},'' in \emph{Proc. ACM SIGCOMM 2015}, vol.~45,
  no.~4, London, UK, Aug. 2015, pp. 269--282.

\bibitem{Vasisht2016NSDI}
D.~Vasisht, S.~Kumar, and D.~Katabi, ``Decimeter-level localization with a
  single {WiFi} access point,'' in \emph{Proc. 13th USENIX Symposium on
  Networked Systems Design and Implementation (NSDI 16)}, Santa Clara, CA, USA,
  Mar. 2016, pp. 165--178.

\bibitem{Wang2016UbiCompRespiration}
H.~Wang, D.~Zhang, J.~Ma, Y.~Wang, Y.~Wang, D.~Wu, T.~Gu, and B.~Xie, ``Human
  respiration detection with commodity {WiFi} devices: Do user location and
  body orientation matter?'' in \emph{Proc. ACM UbiComp 2016}, Heidelberg,
  Germany, Sep. 2016, pp. 25--36.

\bibitem{Wang2016UbiCompGait}
W.~Wang, A.~X. Liu, and M.~Shahzad, ``Gait recognition using {WiFi} signals,''
  in \emph{Proc. ACM UbiComp 2016}, Heidelberg, Germany, Sep. 2016, p.
  363^^e2^^80^^93373.

\bibitem{Qian2017Mobihoc}
K.~Qian, C.~Wu, Z.~Yang, Y.~Liu, and K.~Jamieson, ``Widar: Decimeter-level
  passive tracking via velocity monitoring with commodity {Wi-Fi},'' in
  \emph{Proc. Mobihoc 2017}, Chennai, India, Jul. 2017, pp. 1--10.

\bibitem{Li2021WCOM}
W.~Li, M.~J. Bocus, C.~Tang, R.~J. Piechocki, K.~Woodbridge, and K.~Chetty,
  ``On {CSI} and passive {Wi-Fi} radar for opportunistic physical activity
  recognition,'' \emph{{IEEE} Trans. Wireless Commun.}, vol.~21, no.~1, pp.
  607--620, Jan. 2022.

\bibitem{Molisch2023}
A.~F. Molisch, \emph{Wireless Communications}, 3rd~ed.\hskip 1em plus 0.5em
  minus 0.4em\relax John Wiley \& Sons, 2023.

\bibitem{Bello1963}
P.~Bello, ``Characterization of randomly time-variant linear channels,''
  \emph{IEEE Trans. Commun. Syst.}, vol.~11, no.~4, pp. 360--393, Dec. 1963.

\bibitem{Jiang2022IOT}
Z.~Jiang, T.~H. Luan, X.~Ren, D.~Lv, H.~Hao, J.~Wang, K.~Zhao, W.~Xi, Y.~Xu,
  and R.~Li, ``Eliminating the barriers: Demystifying {Wi-Fi} baseband design
  and introducing the {PicoScenes} {Wi-Fi} sensing platform,'' \emph{{IEEE}
  Internet Things J.}, vol.~9, no.~6, pp. 4476--4496, Aug. 2022.

\bibitem{merker2023sensors}
S.~Merker, S.~Pastel, D.~B^^c3^^bcrger, A.~Schwadtke, and K.~Witte,
  ``Measurement accuracy of the {HTC VIVE Tracker} 3.0 compared to {Vicon}
  system for generating valid positional feedback in virtual reality,''
  \emph{Sensors}, vol.~23, no.~17, pp. 1--17, Aug. 2023.

\bibitem{Jin1995CRLB}
Q.~Jin, K.~M. Wong, and Z.-Q. Luo, ``The estimation of time delay and {Doppler}
  stretch of wideband signals,'' \emph{{IEEE} Trans. Signal Process.}, vol.~43,
  no.~4, pp. 904--916, Apr. 1995.

\bibitem{willis2005bistatic}
N.~J. Willis, \emph{Bistatic Radar}.\hskip 1em plus 0.5em minus 0.4em\relax
  SciTech Publishing, 2005.

\bibitem{Kanda2022CCNCCSI}
T.~Kanda, T.~Sato, H.~Awano, S.~Kondo, and K.~Yamamoto, ``Respiratory rate
  estimation based on {WiFi} frame capture,'' in \emph{Proc. IEEE Consum.
  Commun. Netw. Conf. (CCNC)}, Online, Jan. 2022, pp. 381--384.

\bibitem{Kanda2025Access}
T.~Kanda, S.~Kondo, H.~Shimomura, T.~Sato, H.~Awano, and K.~Yamamoto,
  ``Beamforming feedback-based respiration and heart rate estimation toward
  firmware-agnostic {WiFi} sensing,'' \emph{{IEEE} Access}, vol.~13, pp.
  146\,008--146\,019, Aug. 2025.

\bibitem{Xu2017IMWUT}
Y.~Xu, W.~Yang, J.~Wang, X.~Zhou, H.~Li, and L.~Huang, ``{WiStep}: Device-free
  step counting with {WiFi} signals,'' \emph{Proc. ACM Interactive Mobile
  Wearable Ubiquitous Technol.}, vol.~1, no.~4, pp. 1--23, Dec. 2017.

\end{thebibliography}
\end{document}